\documentclass[aps,preprint,amssymb,12pt,floatfix]{revtex4}
\usepackage{subfigure}
\usepackage{graphicx}
\usepackage{rotating}
\usepackage{color}

\begin{document}
\title{Forced-unfolding and force-quench refolding of RNA hairpins}
\author{Changbong Hyeon$^1$ and D. Thirumalai$^{1,2}$}
\thanks{Corresponding author phone: 301-405-4803; fax: 301-314-9404; thirum@glue.umd.edu}
\affiliation{$^1$Biophysics Program\\
Institute for Physical Science and Technology\\
$^2$Department of Chemistry and Biochemistry\\
University of Maryland, College Park, MD 20742\\
\[
\]
}

\date{\small \today}

\baselineskip = 14pt
\begin{abstract}
Nanomanipulation of individual RNA molecules, using laser optical tweezers, has made it possible to infer 
the major features of their energy landscape. 
Time dependent mechanical unfolding trajectories, measured at a constant stretching force ($f_S$), of 
simple RNA structures (hairpins and three helix junctions) sandwiched 
between RNA/DNA hybrid handles show that they unfold in a reversible all-or-none manner.
In order to provide a molecular interpretation of the experiments we use a general coarse-grained off-lattice Go-like 
model, in which each nucleotide is represented using three interaction sites. 
Using coarse-grained model we have explored forced-unfolding of RNA hairpin as a function of $f_S$ and the loading rate ($r_f$). 
The simulations and theoretical analysis have been done without and with the handles that are explicitly modeled 
by semiflexible polymer chains. 
The mechanisms and time scales for denaturation by temperature jump and mechanical unfolding are vastly different. 
The directed perturbation of the native state by $f_S$ results in a sequential unfolding of the hairpin starting from their 
ends whereas thermal denaturation occurs stochastically. 
From the dependence of the unfolding rates on $r_f$ and $f_S$ we show that the position of the unfolding 
transition state (TS) is not a constant but moves dramatically as either $r_f$ or $f_S$ is changed. 
The TS movements are interpreted by adopting the Hammond postulate for forced-unfolding. 
Forced-unfolding simulations of RNA, with handles attached to the two ends, show that the value of the unfolding force increases (especially at high pulling speeds) 
as the length of the handles increases. 
The pathways for refolding of RNA from stretched initial conformation, upon quenching $f_S$ to the quench force $f_Q$, are 
highly heterogeneous. 
The refolding times, upon force quench, are at least an order of magnitude greater than those obtained by temperature quench. 
The long $f_Q$-dependent refolding times starting from fully stretched states are analyzed using a model that accounts for 
the microscopic steps in the rate limiting step which involves the trans to gauche transitions of the dihedral angles in the GAAA tetraloop. 
The simulations with explicit molecular model for the handles show that the dynamics of 
force-quench refolding is strongly dependent on the interplay of their contour length and the persistence length, and the RNA persistence length. 
Using the generality of our results we also make a number of precise experimentally testable predictions.
\end{abstract}
\maketitle
\newpage

{\bf INTRODUCTION}\\

RNA molecules adopt precisely defined three dimensional structures in order to perform specific functions \cite{DoudnaNature02}. 
To reveal the folding pathways navigated by RNA en route to their native conformations requires exhaustive exploration of the 
complex underlying energy landscape over a wide range of external conditions. 
In recent years, mechanical force has been used to probe the unfolding of a number of RNA molecules \cite{OnoaCOSB04,TinocoARBBS04,BustamanteARB04}.
Force is a novel way of probing regions of the energy landscape that cannot be accessed by conventional methods 
(temperature changes or variations in counterion concentrations). 
In addition, response of RNA to force is relevant in a number of cellular processes such as 
mRNA translocation through the ribosome and the activity of RNA-dependent RNA polymerases. 
Indeed, many dynamical processes are controlled by deformation of biomolecules by mechanical force. 

By exploiting the ability of single molecule laser optical tweezer (LOT) to control the magnitude of the 
applied force Bustamante and 
coworkers have generated mechanical unfolding trajectories for RNA hairpins and \emph{Tetrahymena thermophila} ribozyme \cite{Bustamante2, Bustamante4}. 
The unfolding of the ribozyme shows multiple routes with great heterogeneity in the unfolding pathways \cite{Bustamante4}. 
In their first study \cite{Bustamante2} they showed that simple RNA constructs (P5ab RNA hairpins or a three helix junction) unfold
reversibly at equilibrium. 
From the time traces of the end-to-end distance ($R$) of P5ab, for a number of force values, Liphardt \emph{et. al.} \cite{Bustamante2} showed that the 
hairpins unfold in a two-state manner. The histograms of time dependent 
$R$ (and assuming ergodicity) were used to calculate the free energy difference between the folded and unfolded states. 
Unfolding kinetics as a function of the stretching force $f_S$ was used to identify the position of the transition states \cite{GaubSCI97,Bustamante2,FernandezSCI04}. 
These experiments and subsequent studies have established force as a viable way of quantitatively probing the RNA energy landscape with 
$R$ serving as a suitable reaction coordinate. 

The experiments by Bustamante and coworkers have led to a number of theoretical and 
computational studies using a variety of different methods \cite{HwaBP01,MezardEPJE02,HwaBP03,MarkoEPJE03,RitortBJ05,HyeonPNAS05}. 
These studies have provided additional insights into the mechanical unfolding of RNA hairpins and ribozymes.
In this paper we build on our previous work \cite{HyeonPNAS05} and new theoretical analysis to address a number of questions that pertain to mechanical unfolding of RNA 
hairpins. 
In addition, we also provide the first report on force-quench refolding of RNA hairpins.
The present paper addresses the following major questions:
\begin{description}
\item[1)] Are there differences in the mechanism of thermal and mechanical unfolding? 
We expect these two processes to proceed by different pathways 
because denaturation induced by temperature jump results in a stochastic perturbation of 
the native state while destabilization by force is due to a directed perturbation. 
The molecular model for P5GA gives a microscopic picture of these profound differences. 
\item[2)] For a given sequence does the position of the transition state move in response to changes in the loading rate ($r_f$) or 
the stretching force $f_S$? 
Based on the analysis of experiments over a narrow range of conditions (fixed temperature and loading rate) it has been suggested 
that the location of the sequence-dependent
unfolding transition state (TS) for secondary structure is midway between the folded state while the TS for the ribozyme is close to the native conformation \cite{Bustamante2}. 
Explicit simulations show that the TS moves dramatically especially at high values of $f_S$ and $r_f$. 
As a consequence the dependence of the unfolding rate on $f_S$ deviates from the predictions of the Bell model \cite{BellSCI78}. 
\item[3)] What is the origin of the dramatic differences in refolding by force-quench from stretched conformations and temperature-quench refolding? 
Experiments by Fernandez and Li \cite{FernandezSCI04} on refolding initiated by force-quench on polyubiquitin 
construct suggest similar differences in the refolding time scales. 
For RNA hairpins we show that the incompatibility of the local loop structures in the stretched state and 
the folded conformations lead to extremely long refolding times upon force-quench. 
\item[4)] What are the effects of linker dynamics on forced-unfolding and force-quench refolding of RNA hairpins? 
The manipulation of RNA is done by attaching a handle or linker to the 3' and 5' ends of RNA. 
Linkers in the LOT experiments, that are done under near-equilibrium conditions, are RNA/DNA hybrid handles. 
These are appropriately modeled as worm-like chain (WLC). 
By adopting an explicit polymer model for semiflexible chains we show that, under certain circumstance, 
non-equilibrium response of the handle (which is not relevant for the LOT experiments) can alter the forced-unfolding dynamics of RNA. 
We probe the effects of varying the linker characteristics on the dynamics of folding/unfolding of RNA for the model of the handle-RNA-handle construct. 
In certain range of $r_f$ non-equilibrium effects on the dynamics of linkers can affect the force-extension profiles. 
\end{description}

{\bf METHODS}\\

{\bf Hairpin sequence: }
To probe the dynamics of unfolding and force-quench refolding we have studied in detail the 
22-nucleotide hairpin, P5GA whose solution structure has been determined by NMR (Protein Data Bank (PDB) id:1eor). 
In many respects P5GA is similar to P5ab in the P5abc domain of group I intron \cite{CateScience96}. 
Both these structures have GA mismatches and are characterized by the presence of the 
GAAA tetraloop.
The sequence of P5GA is GGCGAAGUCGAAAGAUGGCGCC \cite{TinocoJMB2000}. \\

{\bf RNA model: }
Because it is difficult to explore, using all atom models of biomolecules in explicit water, unfolding and refolding of RNA hairpins over a wide range of external conditions (temperature ($T$), stretching force ($f_S$), and quench force ($f_Q$)), we have introduced a minimal off-lattice coarse-grained model \cite{HyeonPNAS05} 
(Throughout the paper we use $f$ for referring to mechanical force in general terms while $f_S$ and $f_Q$ have specific meaning).  
In this model each nucleotide is represented by three beads with interaction sites corresponding to the phosphate (P) group, the ribose (S) group, and the base (B) \cite{HyeonPNAS05}.
Thus, the RNA backbone is reduced to the polymeric structure [$(P-S)_n$] with the base that is covalently attached to the ribose center. 
In the minimal model the RNA molecule with N nucleotides corresponds to $3N$ interaction centers.
The secondary structure and the lowest energy structure using the minimal model are shown in Fig.\ref{coarse_RNA}.
\\

{\bf Energy function: }
The total energy of a RNA conformation, that is specified by the coordinates of the 3N sites, 
is written as $V_{T}=V_{BL}+V_{BA}+V_{DIH}+V_{STACK}+V_{NON}+V_{ELEC}$. 
Harmonic potentials are used to enforce structural connectivity and backbone rigidity.
The connectivity between two beads
($P_iS_i$, $S_iP_{i+1}$ and $B_iS_i$) is maintained using 
\begin{equation}
V_{BL}=\sum_{i=1}^{2N-2}\frac{1}{2}k_r\{|\vec{r}_{(SP)_{i+1}}-\vec{r}_{(SP)_i}|-(R_{SP}^o)_i\}^2
+\sum_{i=1}^{N}\frac{1}{2}k_r\{|\vec{r}_{B_i}-\vec{r}_{S_i}|-(R_{BS}^o)_i\}^2
\end{equation}
where $k_r=20$ $kcal/(mol\cdot\AA^2)$, $(R_{SP}^o)_i$ and $(R_{BS}^o)_i$ are the distances between covalently bonded beads in 
PDB structure. 
The notation $(SP)_i$, denotes the $i^{th}$ backbone bead $S$ or $P$. 
The angle $\theta$ formed between
three successive beads ($P_i-S_i-P_{i+1}$ or $S_{i-1}-P_i-S_i$) along sugar-phosphate
backbone is subject to the bond-angle potential,
\begin{equation}
V_{BA}=\sum_{i=1}^{2N-3}\frac{1}{2}k_\theta(\theta_i-\theta_i^o)^2
\end{equation}
where $k_{\theta}=20$ $kcal/(mol\cdot rad^2)$, and $\theta_i^o$ is the value in the PDB structure.\\

{\it Dihedral angle potential :} The dihedral angle potential ($V_{DIH}$) describes 
the ease of rotation around the angle formed between
four successive beads along the sugar-phosphate backbone ($S_{i-1}P_iS_iP_{i+1}$
or
$P_iS_iP_{i+1}S_{i+1}$).
The $i$-th dihedral angle $\phi_i$, which is the angle formed between the two planes defined by four successive beads $i$ to $i+3$, 
is defined by
$cos\phi_i=(\vec{r}_{i+1,i}\times\vec{r}_{i+1,i+2})\cdot(\vec{r}_{i+2,i+1}\times
\vec{r}_{i+2,i+3})$.
In the coarse-grained model
the right-handed chirality of RNA is realized by appropriate choices of the parameters 
in the dihedral potential.
Based on the angles in the PDB structure ($\phi_i^{o}$), one of the three types of
dihedral potentials, trans ($t$, $0<\phi_i^{o} < 2\pi/3$), gauche$(+)$ ($g^+$, $2\pi/3<\phi_i^{o}<4\pi/3$),
gauche($-$) ($g^-$, $4\pi/3<\phi_i^{o}<2\pi$), 
is assigned to each of the four successive beads along the backbone.
The total dihedral potential of the hairpin is 
\begin{eqnarray}
V_{DIH}&=&\sum_{i=1}^{2N-4}[A_{1i}^\eta+B_{1i}^\eta+C_{1i}^\eta\nonumber\\
&+& A_{2i}^\eta cos(\phi_i-\phi_i^{o}+\phi_i^\eta)+B_{2i}^\eta cos3(\phi_i-\phi
_i^{o}+\phi_i^\eta)+C_{2i}^\eta sin(\phi_i-\phi_i^{o}+\phi_i^\eta)]
\label{eqn:DIH}
\end{eqnarray}
where the parameters (in $kcal/mol$) defined for $t$, $g^+$, and $g^-$ are

$A_{1i}=1.0$, $A_{2i}=-1.0$, $B_{1i}=B_{2i}=1.6$, $C_{1i}=2.0$, $C_{2i}=-2.0$ ($\eta=g^+$),

$A_{1i}=1.0$, $A_{2i}=-1.0$, $B_{1i}=B_{2i}=1.6$, $C_{1i}=2.0$, $C_{2i}=2.0$ ($\eta=g^-$),

$A_{1i}=A_{2i}=1.2$, $B_{1i}=B_{2i}=1.2$, $C_{1i}=C_{2i}=0.0$ ($\eta=t$).

\noindent To account for the flexibility in the loop region we reduce the 
dihedral angle barrier by halving the parameter values in $19\leq i\leq 24$.\\

{\it Stacking interactions:}
Simple RNA secondary structures, such as hairpins, are largely stabilized 
by stacking interactions
whose context dependent values are known \cite{WalterPNAS94,MathewsJMB99,DimaJMB05}.
The folded P5GA RNA hairpin is stabilized by nine hydrogen bonds between the base pairs (see Fig.\ref{coarse_RNA}-B)
including two GA mismatch pairs \cite{TinocoJMB2000}.
The stacking interactions that stabilize a hairpin can be written as 
$V_{STACK}=\sum_{i=1}^{n_{max}}V_i$ ($n_{max}=8$ in P5GA).
We assume that the orientation-dependent $V_i$ is 
\begin{eqnarray}
V_i(\{\phi\},\{\psi\},\{r\};T)=\Delta G_i(T)&\times&e^{-\alpha_{st}\{sin^2(\phi_{1i}-\phi_{1i}^{o})+sin^2(\phi_{2i}-\phi_{2i}^{o})+sin^2(\phi_{3i}-\phi_{3i}^{o})+sin^2(\phi_{4i}-\phi_{4i}^{o})\}}\nonumber\\
&\times&e^{-\beta_{st}\{(r_{ij}-r_{1i}^{o})^2+(r_{i+1j-1}-r_{2i}^{o})^2\}}\nonumber\\
&\times&e^{-\gamma_{st}\{sin^2(\psi_{1i}-\psi_{1i}^{o})+sin^2(\psi_{2i}-\psi_{2i}^{o})\}}
\end{eqnarray}
where $\Delta G(T)=\Delta H-T\Delta S$, the bond angles $\{\phi\}$ are
$\phi_{1i}\equiv\angle S_iB_iB_j$, $\phi_{2i}\equiv\angle B_iB_jS_j$,
$\phi_{3i}\equiv\angle S_{i+1}B_{i+1}B_{j-1}$, $\phi_{4i}\equiv\angle B_{i+1}B_{j-1}S_{j-1}$,
the distance between two paired bases $r_{ij}=|B_i-B_j|$, $r_{i+1j-1}=|B_{i+1}-B_{j-1}|$, and
$\psi_{1i}$ and $\psi_{2i}$ are the dihedral angles formed by the four beads
$B_iS_iS_{i+1}B_{i+1}$ and $B_{j-1}S_{j-1}S_jB_j$, respectively.
The superscript $o$ refers to angles and distances in the PDB structure.
The values of $\alpha_{st}$, $\beta_{st}$ and $\gamma_{st}$ are
1.0, 0.3\AA$^{-2}$ and 1.0 respectively.
We take $\Delta H$ and $\Delta S$ from Turner's thermodynamic
data set \cite{MathewsJMB99,WalterPNAS94}.
There are no estimates for GA related stacking interactions.
Nucleotides G and A do not typically form a stable bond and hence GA pairing is considered a mismatch.
We use the energy associated with GU for GA base pair.\\

{\it Nonbonded interactions:} We use the Lennard-Jones
interactions between non-bonded interaction centers to account for the hydrophobicity of the purine/pyrimidine groups. 
The total nonbonded potential is
\begin{equation}
V_{NON}=\sum_{i=1}^{N-1}\sum_{j=i+1}^NV_{B_iB_j}(r)+\sum_{i=1}^N\sum_{m=1}^{2N-1}{'}V_{B_i(SP)_m}(r)+\sum_{m=1}^{2N-4}\sum_{n=m+3}^{2N-1}V_{(SP)_m(SP)_n}(r)
\label{eqn:V_non}
\end{equation}
where $r=|\vec{r}_i-\vec{r}_j|$.
The prime in the second term on the Eq.(\ref{eqn:V_non}) denotes the condition $m\neq2i-1$.
In our model, a native contact exists between two non-covalently bound beads provided they are within
a cut-off distance $r_c$ (=7.0\AA).
Two beads beyond $r_c$ are considered to be non-native.
For a native contact,
\begin{equation}
V_{\xi_i\eta_j}(r)=C_h^{\xi_i\eta_j}[(\frac{r^o_{ij}}{r})^{12}-2(\frac{r^o_{ij}}{r})^6]
\end{equation}
where $r^o_{ij}$ is the distance between beads in PDB structure and $C_h^{\xi_i\eta_j}=1.5$ $kcal/mol$ for all native contact pairs except
for $B_{10}B_{13}$ base pair associated with the formation of the hairpin loop,
for which $C_h^{B_{10}B_{13}}=3.0$ $kcal/mol$.
The additional stability for the base pair associated with loop formation is similar to the
Turner's thermodynamic rule for the free energy gain in the tetraloop region.
For beads beyond $r_c$ the interaction is
\begin{equation}
V_{\xi_i\eta_j}(r)=C_R [(\frac{a}{r})^{12}+(\frac{a}{r})^6]
\end{equation}
with $a=3.4$\AA\ and $C_R=1$ $kcal/mol$.
The value of $C_h^{\xi_i\eta_j}(=1.5kcal/mol)$ has been chosen so that the hairpin undergoes a first order transition from unfolded states. 
Our results are not sensitive to minor variations in $C_h$.\\

{\it Electrostatic interactions:} 
The charges on the phosphate groups are efficiently screened by counterions so that 
in the folded state the destabilizing contribution of the electrostatic potential 
is relatively small. 
However, the nature of the RNA conformation (especially tertiary interactions) can be modulated by changing counterions. 
For simplicity, we assume that
the electrostatic potential between the phosphate groups is pairwise additive
$V_{ELEC}=\sum_{i=1}^{N-1}\sum_{j=i+1}^NV_{P_iP_j}(r)$.
For $V_{P_iP_j}(r)$ we use Debye-H\"{u}ckel potential, 
which accounts for screening by condensed
counterions and hydration effects, and is given by
\begin{equation}
V_{P_iP_j}=\frac{z_{P_i}z_{P_j}e^2}{4\pi\epsilon_0\epsilon_r r}e^{-r/\l_D}
\label{eqn:electrostatic}
\end{equation}
where $z_{P_i}=-1$ is the charge on the phosphate ion, $\epsilon_r=\epsilon/\epsilon_0$
and the Debye length $l_D=\sqrt{\frac{\epsilon_rk_BT}{8\pi k_{elec}e^2I}}$ with $k_{elec}=\frac{1}{4\pi\epsilon_0}=8.99\times 10^{9}JmC^{-2}$.
To calculate the ionic strength $I=1/2\sum_iz_i^2c_i$,
we use the value $c_i=200mM$-$NaCl$ from the header of PDB file \cite{TinocoJMB2000}.
We use $\epsilon_r=10$ in the simulation \cite{MisraPNAS01}.
Because the Debye screening length $\sim\sqrt{T}$ the strength of electrostatic interactions between the phosphate groups are
temperature dependent even when we ignore the variations of $\epsilon$ with $T$.
At room temperature ($T\sim 300$ $K$)
the electrostatic repulsion between the phosphate groups at $r\sim$5.8 \AA,
which is the closest distance between phosphate groups, is $V_{P_iP_{i+1}}\sim 0.5$ $kcal/mol$.
Thus, $V_{ELEC}$ between phosphate groups across the base pairing ($r=16\sim18$ \AA)
is almost negligible. The Debye-H\"{u}ckel interactions is most appropriate for monovalent 
counterions like Na$^+$. 
\\

{\bf Models for linker or ``handles'': }
In laser optical tweezer (LOT) experiments the RNA molecules are attached to the polystyrene 
beads by an RNA/DNA hybrid handles or linker. 
A schematic illustration of the pulling simulations that mimic
the experimental setups for 
the laser optical tweezer (LOT) (Fig.\ref{exp}-A) is shown in Fig.\ref{exp}-B. 
Globally, the principles involved in atomic force microscope (AFM) and LOT for mechanical unfolding of biomolecules 
are essentially the same except for the difference in the effective spring constant.  
The spring constant of the nearly harmonic potential of the optical trap is in range 
$k=0.01-0.1$ $pN/nm$ whereas the cantilever spring in AFM experiments (Fig.\ref{exp}-B) is much stiffer and varies from $1-10$ $pN/nm$. 
To fully characterize the RNA energy landscape it is necessary to explore a wide range of loading rates \cite{EvansNature99}. 
To vary $r_f$ we have used different values of $k$ in the simulations. 

We simulate mechanical unfolding of RNA hairpins either by applying a constant force on the 3'- end with the 5'- end being fixed (unfolding at constant force), 
or by pulling the 3'- end through a combination of linkers and harmonic spring (Fig.\ref{exp}-B) at a constant speed in one direction (unfolding at constant loading rate). 
Comparison of the results allows us to test the role of the linker 
dynamics on the experimental outcome. 
If the linker is sufficiently stiff then it should not affect the dynamics of RNA. 
On the other hand, unfolding at constant loading rate ($r_f\equiv k\times v$, where $v$ is the pulling speed and the stretching force ($f_S$) is computed using 
$f=k\times\delta z$ with $\delta z$ being the displacement of the spring) can be modulated 
either by changing $k$ or $v$.
Instead of $k$ an effective spring constant $k_{eff}$, 
with $k_{eff}^{-1}=k_{eff}^{-1}+k_{linker}^{-1}+k_{mol}^{-1}$,
should be used to compute the loading rate. Typically, $k_{linker}^{-1}$ and $k_{mol}^{-1}$ (where $k_{mol}$ is the stiffness associated with the model) are small and, hence $k_{eff}\approx k$. 
In this setup, the relevant variables are $k$, $v$, and the contour length ($L$) 
of the linker and the effective stiffness of the linker. 
We explore the effects of these factors by probing changes in the force extension curve (FEC). 
The variations in $L$ are explored only using constant loading rate simulations.
Manosas and Ritort \cite{RitortBJ05} used an approximate method to model linker dynamics.

The energy function for the linker molecule is
\begin{equation}
V_L=\sum_{i=1}^{N-1}\frac{k_B}{2}(r_{i,i+1}-b)^2-\sum_{i=1}^{N-2}k_A\hat{r}_{i,i+1}\cdot\hat{r}_{i+1,i+2}
\label{eqn:Linker_Hamiltonian}
\end{equation}
where $r_{i,i+1}$ and $\hat{r}_{i,i+1}$ are the distance and unit vector connecting $i$ and $i+1$ residue, respectively. 
For the bond potential we set $k_B=20$ $kcal/(mol\cdot\AA^2)$ and $b=5$ $\AA$. 
This form of the energy function describes the worm-like chain (WLC) \cite{LeeBJ04} (appropriate for RNA/DNA hybrid handles used by Liphardt \emph{et. al.} \cite{Bustamante2}) when $k_A$ is large. 
The linker becomes more flexible as the parameter describing the bending potential, $k_A$, is reduced. 
Thus, by varying $k_A$ the changes in the entropic elasticity of the linker on RNA hairpin 
dynamics can be examined. 
We use $k_A=80$ $kcal/mol$ or 
$20$ $kcal/mol$ to change  the flexibility of the linker. 
For the purpose of computational efficiency,
we did not include excluded volume interactions between linkers or between the linker and RNA. 
When linkers are under tension the chains do not cross unless 
thermal fluctuations are larger than the energies associated with force. 
To study the linker effect on force extension curves (FEC), we attach two linker polymers each with the contour 
length $L/2$ 
to the ends of the hairpin 
and stretch the molecule using the single pulling speed $0.86\times 10^2$ $\mu m/s$ 
with spring constant $k=0.7$ $pN/nm$. The total linker length is varied from $L=(10-50)$ $nm$. \\

{\bf Simulations: }
We assume that the dynamics of the molecules 
(RNA hairpins and the linkers) can be described by the Langevin equation. 
The system of Langevin equations is integrated as described before \cite{VeitshansFoldDes96,HyeonPNAS05}. 
Using typical values for the mass of a bead in a nucleotide ($B_i$, $S_i$ or $P_i$),
$m=100$ $g/mol\sim160$ $g/mol$,
the average distance between the adjacent beads $a=4.6$ \AA,
the energy scale $\epsilon_h=1\sim2$ $kcal/mol$, the natural time is $\tau_L=(\frac{ma^2}{\epsilon_h})^{1/2}=1.6\sim2.8$ $ps$.
We use $\tau_L=2.0$ $ps$ to convert the simulation times into real times.
To estimate the time scale for thermal and mechanical unfolding dynamics we use a Brownian dynamics algorithm \cite{McCammonJCP78,KlimovFoldDes98}
for which the natural time for the overdamped motion is $\tau_H=\frac{\zeta\epsilon_h}{T}\tau_L$.
We use $\zeta=50\tau_L^{-1}$, which approximately corresponds to friction constant in water, 
in the overdamped limit.
To probe the thermodynamics and kinetics of folding we used a number of physical quantities (end-to-end distance ($R$),
fraction of native contacts ($Q$), the structural overlap function ($\chi$), number of hydrogen bonds $n_{bond}$, etc)
to monitor the structural change in the hairpin.\\

{\bf Computation of free energy profiles: }
We adopted the multiple histogram technique \cite{SwendsenPRL88,KumarJCC1992} 
to compute the thermodynamic averages of all the observables at various values of $T$ and $f$. For example, the thermodynamic average of the fraction of native contact, $Q$, can be obtained at arbitrary values of $T$ and $f$ if the conformational states are well sampled over a range of $T$ and $f$ values. 
The thermodynamic average of $Q$ is given by
\begin{equation}
\langle Q(T,f)\rangle=\frac{\sum_{E,R,Q}Qe^{-(E-fR)/T}\frac{\sum_{k=1}^Kh_k(E,R,Q)}{\sum_{k=1}^K n_ke^{(F_k-(E-f_kR))/T_k}}}{\sum_{E,R,Q}e^{-(E-fR)/T}\frac{\sum_{k=1}^Kh_k(E,R,Q)}{\sum_{k=1}^K n_ke^{(F_k-(E-f_kR))/T_k}}}\equiv\sum_QQP[Q(T,f)]
\label{eq:WHAM}
\end{equation}
where $K$ is the number of histograms, 
$h_k(E,R,Q)$ is the number of states between $E$ and $E+\delta E$, $R$ and $R+\delta R$, $Q$ and $Q+\delta Q$ in
the $k$-th histogram, $n_k=\sum_{E,R,Q}h_k(E,R,Q)$,
$T_k$ and $f_k$ are the temperature and the force in the simulations used to generate the 
$k-$th histogram, respectively. 
The free energy, $F_k$, that is calculated self-consistently, satisfies 
\begin{equation}
e^{-F_r/T_r}=\sum_{E,R,Q}e^{-(E-f_rR)/T_r}\frac{\sum_{k=1}^Kh_k(E,R,Q)}{\sum_{k=1}^Kn_ke^{(F_k-(E-f_kR))/T_k}}.
\label{eq:WHAM2}
\end{equation}
Using the low friction Langevin dynamics, 
we sampled the conformational states in the ($T$,$f$) in the range 
$\{0$ $K<T<500$ $K,f=0.0$ $pN\}$ and $\{0.0$ $pN<f<20.0$ $pN,T=305$ $K\}$.
Exhaustive samplings around the transition regions at $\{305$ $K\leq T\leq 356$ $K$, $f=0.0$ $pN\}$ and
$\{5.0$ $pN\leq f\leq7.0$ $pN$, $T=305$ $K\}$ is required to obtain reliable estimates of the thermodynamic quantities. 
The free energy profile, with $Q$ as an order parameter, is given by 
\begin{equation}
F[Q(T,f)]=F_o(T,f)-k_BT\log{P[Q(T,f)]}.
\label{eq:freeprofile}
\end{equation}
where $F_o(T,f)=-k_BT\log{Z(T,f)}$, 
$Z(T,f)=\sum_{E,R,Q}e^{-(E-fR)/T}\frac{\sum_{k=1}^Kh_k(E,R,Q)}{\sum_{k=1}^K n_ke^{(F_k-(E-f_kR))/T_k}}$ and  $P[Q(T,f)]$ is defined in Eq.(\ref{eq:WHAM}).
The free energy profile $F(R)$ with $R$ as a reaction coordinate can be  obtained using a similar expression. \\

{\bf RESULTS}\\

{\bf Mechanisms of thermal denaturation and forced-unfolding are different: }
We had previously reported \cite{HyeonPNAS05} the thermodynamic characteristics of the P5GA hairpin as a function of $T$ and $f$. 
The native structure of the hairpin, that was determined using a combination of multiple slow 
cooling, simulated annealing, and steepest descent quench, yielded conformations whose 
average root mean square deviation (RMSD) with respect to the PDB structure is about 0.1 \AA\ \cite{HyeonPNAS05}. The use of Go-like model leads to a small value of RMSD. 
From the equilibrium ($T$,$f$) phase diagram in \cite{HyeonPNAS05} the melting 
temperature $T_m\approx 341$ $K$ at $f_S=0$. 
Just as in thermal denaturation at $T_m$ at zero $f$ 
the hairpin unfolds by a weak first order transition at an equilibrium critical force $f_c$. Above $f_c$,  
which is a temperature-dependent, the folded state is unstable. 

To monitor the pathways explored in the thermal denaturation 
we initially equilibrated the conformations at $T=100$ $K$ at which the hairpin is stable. 
Thermal unfolding ($f=0$) was initiated by a temperature jump to $T=346$ $K>T_m$. 
Similarly, forced-unfolding is induced by applying a constant force $f_S=42$ $pN$ to 
thermally equilibrated initial conformation at $T=254$ $K$ \cite{HyeonPNAS05}. 
The value of $f_S=42$ $pN$ far exceeds $f_c=15$ $pN$ at $T=254$ $K$. 
Upon thermal denaturation, 
the nine bonds fluctuate (in time) stochastically in a manner that is 
independent of one another until the hairpin melts (Fig.\ref{unfoldpath}-A). 
Forced-unfolding, on the other hand, occurs in a directed manner. 
Mechanical unfolding occurs by sequential unzipping with force unfolding the bond, from the ends of the hairpin (beginning at bond 1) to the loop (Fig.\ref{unfoldpath}-B). 

The differences in the folding pathways are also mirrored in the free energy profiles. 
Assuming that $Q$ is an adequate reaction coordinate for thermal unfolding \cite{OnuchicPNAS98} we find, by comparing $F(Q)$ at $T=100$ $K$ and $T=346$ $K$, that thermal melting occurs by crossing a barrier.
The native basin of attraction (NBA) at $Q=0.9$ at $T=100$ $K$ is unstable at the higher temperature and the new equilibrium at $Q\sim 0.2$ is reached in an apparent two state manner (Fig.\ref{unfoldpath}-C). 
Upon ``directed'' mechanical unfolding the free energy profile $F(R)$ is tilted from the NBA at $R\approx 1.5$ $nm$ to $R\approx 12$ $nm$ at which the stretched states are 
favored at $f=42$ $pN$. 
The forced-unfolding transition also occurs abruptly once the activation barrier at 
$R\approx 1.5$ $nm$ (close to the folded state) is crossed 
(Fig.\ref{unfoldpath}-D). \\

{\bf Free energy profiles and transition state (TS) movements: }
Based on the proximity of the average transition state location, $\Delta x_F^{TS}$, it has been suggested that folded states of RNA \cite{Bustamante4} 
and proteins \cite{GaubSCI97} are brittle. 
If the experiments are performed by stretching at a constant loading rate then $\Delta x_F^{TS}$ 
is calculated using $f^*\sim k_BT/\Delta x_F^{TS}\log{r_f}$ \cite{Reif}
where $f^*$ is the most probable unfolding force and $r_f$, the loading rate, is $r_f=df/dt=kv$. 
Substantial curvatures in the dependence of $f^*$ on $\log{r_f}$ ([$f^*$,$\log{r_f}$] plot) have been observed especially if $r_f$ 
is varied over a wide range \cite{EvansNature99}. 
Similarly, in constant force unfolding experiments $\Delta x_F^{TS}$ is obtained 
from the Bell equation \cite{BellSCI78} that relates the unfolding rate to the applied force, 
$\log{k_U}=\log{k^o_U}+f\Delta x_F^{TS}/k_BT$ where $k^o_U$ is the unfolding rate in the absence of force. 
In the presence of curvature in the [$f^*$,$\log{r_f}$] plots or when the Bell relationship is violated \cite{HummerBP03} 
it is difficult to extract meaningful values of $k^o_U$ or $\Delta x_F^{TS}$ by a simple linear extrapolation. 
By carefully examining the origin of curvature in the [$f^*$,$\log{r_f}$] plot or in $k_U$ as we show that in the 
unfolding of hairpin the observed non-linearities are due to movements in the transition state ensemble i.e., $\Delta x_F^{TS}$ depends on $f_S$ and $r_f$.\\

{\it Unfolding at constant loading rate: }
We performed forced unfolding simulations by varying 
both the pulling speed $v$ and the spring constant $k$ so that a broad range of loading rates can be covered. 
The unfolding forces 
at which all the hydrogen bonds are ruptured are broadly distributed with the 
average and the dispersion that increase with growing loading rates 
(Fig.\ref{pullsummary}-A).
The plot of $f^*$ as a function of $\log{r_f}$ (Fig.\ref{pullsummary}-B) shows 
marked departure from linearity. 
The slope of the plot ([$f^*$,$\log{r_f}$]) increases sharply as $r_f$ increases. 
There are two possible reasons for the increasing tangent. 
One is the reduction of $\Delta x^{TS}_F$, which would lead to an increase in the slope ($k_BT/\Delta x_F^{TS}$) 
of $f^*$ vs $\log{r_f}$.
The other is the increase of curvature at the transition state region, i.e., barrier top of 
free energy landscape. 
Regardless of the precise reason, it is clear that the standard way of estimating $\Delta x_F^{TS}$ using $f^*$ at large loading rates 
results in a very small value of $\Delta x_F^{TS}$. 
We estimate $\Delta x_F^{TS}$ from [$f^*$,$\log{r_f}$] plot to be 4 \AA\ at $r_f\approx 10^5$ $pN/s$. 
The estimated value of $\Delta x_F^{TS}$ is unphysical because 4 \AA\ is less than the average distance between neighboring $P$ atoms. 
The minimum pulling speed used in our simulations is nearly five orders of magnitude greater than in experiments. 
The use of large loading rate results in small values of $\Delta x_F^{TS}$.
If the simulations can be performed at small values of $r_f$ we expect the slope of $f^*$ vs $\log{r_f}$ to decrease, 
which would then give rise to physically reasonable values of $\Delta x_F^{TS}$ at low $r_f$. 
Our simulations suggest that the curvature in the plot of $f^*$ as a function of $\log{r_f}$ is due to the dependence of $\Delta x_F^{TS}$ on $r_f$ and not due 
to the presence of multiple transition states \cite{EvansNature99}. As a result, extrapolation to low $r_f$ values can give erroneous results (Fig.\ref{pullsummary}-B).\\

{\it Unfolding at constant force: }
To monitor the transition state movements we performed a number of unfolding simulations by applying $f_S>20$ $pN$ at $T=290$ $K$. 
The unfolding rates are too slow at $f_S<20$ $pN$ to be simulated. 
Nevertheless, the simulations give strong evidence for force-dependent movement of $\Delta x_F^{TS}$.  
For a number of values of $f_S$ in the range $20$ $pN<f_S<150$ $pN$ we computed the distribution of first passage times for unfolding. 
The first passage time for the $i$-th molecule is reached if $R$ becomes $R=5$ $nm$ for the first time. 
From the distribution of first passage times (for about (50-100) molecules at each $f_S$) 
we calculated the mean unfolding time. 
Just as for unfolding at constant $r_f$ the dependence of $\log{\tau_U}$ on 
$f_S$ shows curvature (Fig.\ref{unfoldrefold}-A), and hence deviates from the often used Bell model \cite{HummerBP03}. 
By fitting $\tau_U$ to the Bell formula ($\tau_{U}=\tau^o_Ue^{-f_S\Delta x_F^{TS}/k_BT}$), \emph{over a narrow range of $f_S$} 
we obtain $\Delta x_F^{TS}\approx 4$\AA\ which is too small to be physically meaningful at $f_S= 0$.

Insights into the shift of $\Delta x_F^{TS}$ as $f_S$ increases can be gleaned from the equilibrium  
force-dependent free energy $F(R)$ as a function of $R$.
The one-dimensional free energy profiles $F(R)$ show significant movements in $\Delta x_F^{TS}$ as $f_S$ changes (Fig.\ref{unfoldrefold}-C). 
As $f_S$ increases $\Delta x_F^{TS}$ decreases sharply, which implies that the unfolding TS is 
close to the folded state. 
At smaller values of $f_S$ the TS moves away from the native state. 
At the midpoint of the transition $\Delta x_F^{TS}\approx 5.5$ $nm$ which is about half-way to the native state. 
The result for $\Delta x_F^{TS}/R_U\approx 1/2$ ($R_U$ is the average value of $R$ in the unfolded state) is in accord with experiments \cite{Bustamante2}
which were done at forces that are not too far from the equilibrium unfolding force. 
Although $\Delta x_F^{TS}$ is dependent on the RNA sequence it is likely that, for simple hairpins $\Delta x_F^{TS}/R_U\approx 1/2$.
The prediction that $\Delta x_F^{TS}$ \emph{is dependent on $f_S$} is amenable to experimental test. \\

{\bf Force-quench refolding times depend (approximately) exponentially on $f_Q$: }
One of the great advantages of force quench refolding experiments \cite{FernandezSCI04} is that the ensemble of conformations with a predetermined value of $R$ can be prepared by suitably 
adjusting the value of $f_S$ \cite{FernandezSCI04}. 
Because force-quench refolding can be initiated from conformations with arbitrary values of $R$, 
regions of the energy landscape that are completely inaccessible in conventional 
experiments can be probed.
In order to initiate refolding by force quench we generated extended conformations with $R=13.5$ $nm$, using $f_S=90$ $pN$ at $T=290$ $K$.
Subsequently, we reduced the force to $f_Q$ in the 
range $0.5$ $pN<f_Q<4$ $pN$. 
For these values of $f_Q$ the hairpin conformation is preferentially populated at equilibrium. 
The probability $P_U(t)$ that the RNA hairpin remains unfolded at six $f_Q$ values decays non-exponentially (Fig.\ref{lagtimeanalysis}). 
The mean refolding times $\tau_F(f_Q)$, upon force quench, that are computed using $P_U(t)$ 
($\tau_F=\int_0^{\infty}tP_U(t)dt$), 
show that \emph{in the range} $0.5$ $pN<f_Q<4$ $pN$ (Fig.\ref{unfoldrefold}-B)
\begin{equation}
\tau_F(f_Q)=\tau_F^o\exp{(f_Q\Delta x_U^{TS}/k_BT)}
\label{eqn:refold_tau}
\end{equation}
where $\Delta x_U^{TS}$ is the distance from the unfolded basin of attraction to the 
refolding transition state, and $\tau_F^o$ is the refolding time in the absence of 
force. 
Linear regression gives $\tau_F^o\approx 290$ $\mu s$ and $\Delta x_U^{TS}\approx 1$ $nm$. 
The value of $\Delta x_U^{TS}$, which is obtained from kinetic simulations, is in accord with the location of $\Delta x_U^{TS}$ obtained directly from the equilibrium 
free energy profiles $F(R)$ (Fig.\ref{unfoldrefold}-C).
In the $0.5$ $pN<f_Q<4$ $pN$ range the distance from UBA to the TS is approximately 1 $nm$ and is a constant. 
Thus, kinetic and thermodynamic data lead to a consistent picture of force-quench refolding. 

If the simulations are done with $f_Q\equiv 0$ $pN$ then we find that $\tau_F(f_Q=0)\approx 191$ $\mu s$ (Fig.\ref{unfoldrefold}-B) 
which differs from $\tau_F^o$ obtained using Eq.(\ref{eqn:refold_tau}).
When $f_Q$ is set to zero, the 3' end fluctuates whereas when $f_Q\neq 0$ the 3' end remains fixed. 
The rate limiting step in the hairpin formation is the trans$\rightarrow$gauche transitions in the dihedral angles in the GAAA tetraloop region (see below). 
When one end is free to fluctuate, as is the case when $f_Q=0$, the trans$\rightarrow$gauche occurs more rapidly than $f_Q\neq 0$. 
The difference between $\tau_F(f_Q=0)$ and $\tau_F^o$ is, perhaps, 
related to the restriction in the conformational search among the compact structures which occurs when one end of the chain is fixed. 
\\ 

{\bf Metastable intermediates lead to a lag phase in the refolding kinetics: }
The presence of long lived conformations (see below), with several incorrect dihedral 
angles in the GAAA tetraloop, is also reflected in the refolding kinetics as monitored by $P_U(t)$. 
If there are parallel routes to the folded state then $P_U(t)$ can 
be described using a sum of exponentials. 
The lag kinetics, which is more pronounced as $f_Q$ increases (see especially $f_Q=4$ $pN$ in Fig.\ref{lagtimeanalysis}) can be rationalized using a kinetic scheme 
$S \stackrel{\tau_1}\longrightarrow I\stackrel{\tau_2}\rightarrow F$
where $S$ is the initial stretched state, $I$ is the intermediate state, and 
$F$ is the folded hairpin. 
Setting $P_U(t)\equiv P_S(t)+P_I(t)=1-P_F(t)$, we obtain 
\begin{equation}
P_U(t)=\frac{1}{\tau_2-\tau_1}\left(\tau_2 e^{-t/\tau_2}-\tau_1 e^{-t/\tau_1}\right). 
\label{eqn:consecutive_process}
\end{equation}
From Fig.\ref{lagtimeanalysis}, which shows the fits of the simulated $P_U(t)$ to Eq.(\ref{eqn:consecutive_process}), 
we obtain the parameters ($\tau_1$, $\tau_2$) at 
each $f_Q$ (see caption to Fig.\ref{lagtimeanalysis} for the values). 
If folding is initiated by temperature-quench $\tau_1\ll \tau_2$ so that $P_U(t)\sim e^{-t/\tau_2}$. 
Explicit simulations show that thermal refolding occurs in a two-state manner (data not shown). 
In force-quench refolding both $\tau_1$ and $\tau_2$ increase as $f_Q$ increases and $\tau_1/\tau_2\sim O(1)$ at the higher values of $f_Q$. 
There are considerable variations in the structures of the metastable intermediate depending on $f_Q$. 
The variations in the conformations are due to the differences in the number of incorrect or non-native dihedral angles.
As a consequence there are multiple steps in force quench refolding in contrast to forced-unfolding which occurs in an all-or-none manner. \\

{\bf Trans$\rightarrow$gauche transitions in the GAAA tetraloop dihedral angles lead to long refolding times: }
It is of interest to compare the refolding times obtained from stretched ensemble ($\tau_F(f_Q)$) and 
the refolding time ($\tau_F(T)$) from thermally denatured ensemble. 
In a previous paper \cite{HyeonPNAS05}, we showed that $\tau_F(f_Q=0)=15\tau_F(T)$ (Fig.\ref{unfoldrefold}-B).
The large difference in refolding times may be because the initial conditions 
from which folding commences are vastly different upon force and temperature quench \cite{FernandezSCI04_2,HyeonPNAS05}. 
The fully stretched conformations, with $R=13.5$ $nm$, are very unlikely to occur in an equilibrated ensemble even at elevated temperatures. 
The canonical distribution of thermally denatured conformations 
even at $T=1500$ $K$ ($\gg T_F$) shows that the probability of populating conformations with $R=13.5$ $nm$ (Fig.\ref{Ext_refold}-A) is practically zero.
Thus, folding from thermally denatured ensemble starts from relatively compact conformations. In contrast, the initial condition for force-quench refolding can begin (as in our simulations) from fully stretched conformations upon force-quench. 
Both $R$ and the radius of gyration ($R_g$) undergo substantial changes en route to the NBA. 
Indeed, the refolding trajectories from extended conformations exhibit broad 
fluctuations in $R(t)$ in the order of (25-75\AA) for long time periods, followed by a cooperative reduction 
in $R$ at the final stage (Fig.\ref{Ext_refold}-B). 

The long refolding times upon force-quench starting from fully stretched conformations may be generic for folding of globular proteins as well. 
Recent experiments on force-quench refolding of poly-ubiquitin (Ub) 
\cite{FernandezSCI04} show that $R(t)$ for proteins exhibit behavior similar to that shown in Fig.\ref{Ext_refold}-B. 
The resulting $f_Q$-dependent refolding times for poly-Ub ($0.1-10$ $sec$) are considerably larger compared to $\tau_F$($\approx 5$ $msec$ \cite{RoderBC93,SosnickJMB02}) in the absence of force. 
Because the collapse of a single ubiquitin molecule in solution occurs 
in less than a millisecond, the origin of the long refolding times has drawn considerable attention \cite{SosnickSCI04,FernandezSCI04_2}.
The microscopic model considered here for RNA can be used to shed light on the origin of the generic long 
refolding times upon force quench. 

From the phase diagram (see Fig.(2) in \cite{HyeonPNAS05}) 
it is clear that the routes navigated by RNA hairpin upon thermal and force quench 
have to be distinct.
While the distinct initial conditions do not affect the native state stability (as long as the final values of $T$ and 
$f_Q$ are the same) they can profoundly alter the rates and pathways of folding. 
The major reason for the long force-quench refolding times in RNA hairpins 
is that in the initial stretched state there is a severe distortion (compared to its value in the native state) 
in one of the dihedral angles. 
The 19-th dihedral angle (found in the GAAA loop region) along the sugar-phosphate 
backbone is in $g^+$ conformation in the native state while in the initial stretched conformation it is in the $t$ state (Fig.\ref{dihedral}-A). 
Thus, if all the molecules are in the fully stretched conformation then the $19^{th}$ dihedral angle in 
each of them has to, during the force-quench refolding process, undergo the $t\rightarrow g^+$ transition in the GAAA tetraloop region in order to fold. 
The enthalpic barrier associated with the $t\rightarrow g^+$ transition is about $1k_BT$ (Fig.\ref{dihedral}-B). 
However, this transition is coupled to the dynamics in the other degrees of freedom and 
such a cooperative event (see Fig.\ref{Ext_refold}-B for examples of trajectories) makes the effective free energy barrier even higher. 
Although significant fluctuations are found in thermally denatured ensemble at $T=500$ $K$, 
they are not large enough to produce non-native dihedral angles in the GAAA tetraloop.
The dihedral angles in thermally denatured conformations do not deviate significantly from their values 
in the native conformation (Fig.\ref{dihedral}-C).
In contrast upon fully stretching P5GA dihedral angles in the GAAA tetraloop adopt non-native values (Fig.\ref{dihedral}-D).

The time scale for the $t\rightarrow g^+$ transitions can be inferred from the individual trajectories.
Typically, there are large  fluctuations due to $g^+\leftrightarrow t\leftrightarrow g^-$ transitions 
in the dihedral angles in the flexible loop region ($i=19-24$). 
For the trajectory in Fig.\ref{evolutionDIH}-B we observe
the persistence of incorrect dihedral angle in the loop region 
for $t\sim 300$ $\mu s$. 
At $t>300$ $\mu s$ the native-like dihedral angles form. 
Subsequently, the formation and propagation of interaction stabilizing 
the native RNA hairpin takes place. 
These dynamical transitions are clearly observed in Figs.\ref{evolutionDIH}-B and \ref{evolutionDIH}-C.
The observed mechanism is reminiscent of a nucleation process. 
We conclude that the formation of the flexible loop with all the dihedral angles achieving near native values 
is the rate limiting step in the refolding kinetics of RNA hairpins upon force-quench starting from fully stretched state. 
It should be stressed that the rate limiting step for thermal refolding of P5GA or force-quench refolding starting from partially stretched conformations (see below) 
is different. 
The observation that the zipping of the hairpin takes place upon synchronous 
formation of all the native-like dihedral angles suggests the presence of a high entropic barrier. 
The crossing of the entropic barrier results in slow refolding if P5GA is fully stretched. \\

{\bf Linker effects on RNA force-extension curves :}
In LOT experiments the force-extension curves (FECs) are measured for the handle(H)-RNA-handle(H) construct \cite{Bustamante2, Bustamante4}. 
To unambiguously extract the FEC for RNA alone (from the measured FEC for H-RNA-H construct) the properties of the handle, 
namely, the contour length $L_H$ and the persistence length $l_p^H$ cannot be chosen arbitrarily. 
In the experiments by Liphardt \emph{et. al.} $L_H=320$ $nm$ and $l_p^H=50 $ $nm$ for the RNA/DNA hybrid handle. 
We expect that both $\lambda=l_p^H/l_p^{RNA}$ and $l_p^H$ will affect the FEC curves of the object of interest, 
namely, the RNA molecule. 
To discern the signature for the force-induced transition in the RNA hairpin alone from the FEC for H-RNA-H construct
$\lambda$ has to be large. 
If we assume $l_p^{RNA}\approx 1$ $nm$ 
then the experimental value of $\lambda\approx 50$ which is large enough to extract the transitions in the RNA hairpin. 
If $\lambda\approx 1$ then the entropic fluctuations in the handle can mask the signals in RNA \cite{HummerSCI05}. 
Similarly the end-to-end distance fluctuation in the handle, $\delta R$, should be smaller then the extension in RNA. 
Because $\delta R$ grows with $L_H$ (see below) it follows that if very long $L_H$ is used even with $\lambda\gg 1$ the signal from RNA can be masked. 
The square of the fluctuations in the end-to-end distance $\delta R$ of the linker is given by 
$(\delta R)^2=\partial\langle R\rangle/\partial(\beta f_S) $ where $\langle R\rangle$ is the mean end-to-end distance. 
For WLC, that describes the linkers, we expect $\langle R\rangle\sim (2L_Hl_p)^{1/2}$ for small $f_S$ and $\langle R\rangle\sim L_H$ for large $f_S$ 
provided $L_H$ is long. 
Thus, 
\begin{equation}
\delta R\sim\left\{ \begin{array}{ll}
     (L_Hl^H_p)^{1/4}(k_BT/f_S)^{1/2}& \mbox{$(f_S\lesssim k_BT)$}\\
     L_H^{1/2}(k_BT/f_S)^{1/2}& \mbox{$(f_S\gg k_BT)$}\end{array}\right.
\label{eqn:fluctuation}
\end{equation}
The mean fluctuation in the extension of the spring is, using equipartition theorem, given by 
\begin{equation}
\delta x\sim\sqrt{\frac{k_BT}{k}}.
\end{equation}
In order for the signal from RNA to be easily discerned from the experimentally measured FEC,
the expansion of end-to-end distance of the molecule at transition should be larger than 
$\delta R$ and $\delta x$. 
Since $\delta R$ grows sublinearly with the linker length (Eq.(\ref{eqn:fluctuation}))
the attachment of large linker polymer can mask the transition signal. 
These arguments show the characteristics of the linker can obscure the signals from RNA. 

The FECs in the simulations can also be affected by non-equilibrium effects due to the linker dynamics. 
In the handle(H)-RNA-handle(H) construct considered here the force exerted at one of the linker depends on the angle between $f_S$ and the end of the linker. 
The initial event in force transmission along the contour of the H-RNA-H construct is the alignment of the molecule 
along the force direction. 
The characteristic time for force to reach RNA so that unfolding can occur is $f_c/r_f$ where $f_c$ is the critical force. Non-equilibrium effects due to 
linker dynamics become relevant if $\tau_R$, the time scale for alignment of H-RNA-H along the force direction, $\tau_R>f_c/r_f$. 
This condition is not relevant in experiments which are conducted at small values of $r_f$. 
However, it is important to consider non-equilibrium effects in simulations which are performed at high $r_f$ values. 
The characteristic time depends on $L_H$ and $l_p^H$. 

We validate these arguments by obtaining FEC for H-P5GA-H by varying $L_H$ and $\lambda=l_p^H/l_p^{RNA}$. 
Using the WLC for the linker we calculated FEC where the $L_H$ is varied from $(10-50) $ $nm$. 
In order to observe rapid unfolding we have carried out our simulation at the pulling speed $v=0.86\times 10^2$ $\mu m/sec$ and $k=0.7$ $pN/nm$. 
Under these conditions non-equilibrium effects are relevant for the linker \cite{LeeBJ04} which is not the case in experiments. 
The FECs show clearly a plateau in the range $20$ $pN<f_S<40$ $pN$ which corresponds to the two-state hairpin opening (Fig.\ref{linker_anal}). 
For the experimentally relevant plot (Fig.\ref{linker_anal}-A) that shows FEC for H-P5GA-H the transition plateau is present at all values of $L_H$. 
However, when $L_H$ reaches $50$ $nm$ the signal from P5GA is masked.
The FEC for P5GA alone (Fig.\ref{linker_anal}-B) shows modest increase in the unfolding force as $L_H$ increases. 
Similarly, we find the value of unfolding force also increases as the linker flexibility increases. 
These observations are due to non-equilibrium effects on the linker dynamics because of the relatively large values of $r_f$ used in the simulations. 

Our simulations show that, at high loading rates, the length of the handle is also important. 
This issue is not relevant in experiments in which loading rates are much smaller. 
However, they become important in interpreting simulation results. 
In our case $r_f(=kv)$ is $6\times 10^4$ times that used in experiments. 
At such high $r_f$ non-equilibrium effects control the linker dynamics \cite{LeeBJ04}.
Thus, to extract unfolding signatures from RNA alone it is necessary to use high value of $\lambda$ and relatively short values of $L$. 
In other words FEC, when $r_f$ is varied, may be a complicated function of $l_p^H/l_p^{RNA}$ and $L_H/l_p^H$. 
\\

{\bf Force-quench refolding of P5GA with attached linkers:}
It is convenient to monitor force-quench refolding of RNA alone using simulations. 
A similar experiment can only be performed by attaching handles to RNA. 
In such an experiment, which has not yet been done (however, see Note added), the H-RNA-H would be stretched by a stretching force $f_S>f_c$
so that RNA unfolds. 
By a feedback mechanism the force is quenched to $f_Q\neq0$. 
We have simulated this situation for the H-P5GA-H construct with $L_H=15$ $nm$, $l_p^H=30$ $nm$ for each handle.
We chose very stiff handles ($L_H<l_p^H$) so that the dynamics of RNA can be easily monitored. 
The end-to-end distance of the H-P5GA-H system as a function of $t$ with $f_S=90$ $pN$ and $f_Q=2$ $pN$ shows a rapid decrease 
from $R_{sys}=44$ $nm$ to about $R_{sys}=37$ $nm$ in less than about $100$ $\mu s$ (Fig.\ref{refoldlinker}).
The use of large values of $f_S$ will not affect the results qualitatively.
The value of $R_{sys}$ fluctuates around $36$ $nm$ for a prolonged period and eventually $R_{sys}$ attains its equilibrium value around $30$ $nm$. 

Upon decomposing $R_{sys}$ into contributions from P5GA and the handle we find that the major changes in $R_{sys}$ occur 
\emph{when RNA undergoes the folding transition} (compare the top and middle panels in Fig.\ref{refoldlinker}). 
The time dependence of $R_H$, which monitors only the dynamics of the linker, shows that with $f_Q=2$ $pN$ 
after the initial relaxation $R_H$ fluctuates around its equilibrium value 
(bottom panel in Fig.\ref{refoldlinker}). 

From these simulations and others for different $f_Q$ values we can make a few general comments that are 
relevant for experiments. 
(a) As long as $\lambda(=l_p^H/l_p^{RNA})$ is large enough the qualitative aspect of RNA folding can be obtained from the dynamics of $R_{sys}$ alone. 
However, if $L_H/l_p^H\gg 1$ then large transverse fluctuations of the linker can interfere with the signal from RNA molecule. 
(b) To obtain quantitative results for the dynamics of RNA (i.e., $R_M$ as a function of $t$) the dynamics of the handle 
upon force relaxation has to be described accurately. 
Upon $f_S\rightarrow f_Q$ quench the dynamics of the handle cannot be described using Langevin equation using the 
equilibrium force. 
Instead, the relaxation behavior must be determined by solving the Langevin equation for the WLC energy function \cite{Bohbot-RavivPRL04}
which is subject to 
$f_S\rightarrow f_Q$ quench. \\

{\bf DISCUSSIONS}\\

{\bf Transition state movement and Hammond postulate for force: }
Hammond postulate \cite{HammondJACS53,LefflerSCI53} 
is widely used to qualitatively predict the nature of 
transition state in the chemical reactions of organic molecules.
In recent years a number of protein folding experiments have been interpreted using generalization of the Hammond 
postulate \cite{FershtBook}.
The Hammond postulate states that \emph{if a transition state and an unstable intermediate, 
occur consecutively during a reaction process and have nearly the same energy, their interconversion will involve only a 
small reorganization of the molecular structure} \cite{HammondJACS53}. 
In the context of RNA folding, the Hammond postulate suggests that  
\emph{the position of the transition state along the reaction coordinate is shifted towards the destabilized state, either folded or unfolded, depending on the nature of perturbation.}
The Hammond behavior is most vividly seen in the free energy profiles $F(R)$ (Fig.\ref{unfoldrefold}-C) that pictorially describe mechanical unfolding of RNA hairpins. 
As $f$ is increased the unfolded state is preferentially stabilized. 
From the Hammond postulate we would infer that the major TS should be more native-like as $f$ increases. 
The force-dependent $F(R)$ as a function of $R$ indeed confirms (Fig.\ref{unfoldrefold}-C) that as $f$ increases $\Delta x_F^{TS}$ becomes closer to the native folded 
hairpin conformation. 

RNA hairpins also denature upon heating. 
To ascertain the variation in the location of the TS as temperature is changed we have calculated the free energy $F(Q)$ as a function of $Q$ 
at several values of $T$ at $f=0$ (Fig.\ref{Hammond}-A). 
Although the location of the TS follows Hammond behavior there is \emph{very little change} in the TS ensemble over the temperature range examined. 
Thus, the changes in the TS are very dramatic when unfolding is induced by force compared to thermal denaturation. 

Hammond behavior can be quantified using the Leffler's proportionality constant $\alpha_x$
which measures the energetic sensitivity of the transition state relative to 
the native states when the population shift is induced by a perturbation $x$ \cite{LefflerSCI53,KiefhaberJMB03_2}.
For mechanical unfolding ($x=f_S$)
\begin{equation}
\alpha_f=\frac{\partial\Delta F^{\ddagger}(R)/\partial f_S}{\partial\Delta F_{UF}(R)/\partial f_S}=\frac{\Delta x_{F}^{TS}}{\Delta x_{UF}}.
\label{eqn:Lefler}
\end{equation}
Using the free energy profile we computed $\alpha_f$ as a function of $\Delta F_{UF}$ or $f$ (Fig.\ref{Hammond}-B). 
The shift in the transition state is, quantified by $\alpha_f$ in the range $0\leq\alpha_f\leq 1$, and the 
shift rate (or self-interaction parameter, $p_f\equiv\partial\alpha_f/\partial\Delta F_{UF}$ \cite{KiefhaberJMB03_2}) has maximum in the force range $4<f_S<10$ $pN$. 
As $\Delta F_{UF}$ decreases (the UBA is stabilized with respect to the NBA) $\alpha_f$ decreases, 
which implies that the TS becomes increasingly native-like (Fig.\ref{Hammond}-B).
The inset in Fig.\ref{Hammond}-B  
shows dramatically the changes in $\alpha_f$ with respect to $f_S$. 
The largest changes in $\alpha_f$ occurs as $f_S$ approaches the $T$-dependent ($T=290$ $K$) $f_c\approx 7$ $pN$. 
A similar plot of $\alpha_T$ as a function of $T$ shows practically no change in $\alpha_T$. 
From this analysis we conclude that the nature of the transition state ensemble is different in 
mechanical unfolding and thermal denaturation.
 
The transition state movement with force is very sensitive to the shape of 
the barrier in the vicinity of the transition state.  
The free energy profile near the barrier ($x\sim x_{ts}$) can 
be expanded as $F(x)\sim F(x_{ts})-\frac{1}{2}F''(x_{ts})(x-x_{ts})^2+\cdots$. 
Upon application of the stretching force $F(x)$ is tilted by $-f_S\cdot x$. 
The new barrier position ($x_{ts}^{NEW}$) is at  
$x_{ts}^{NEW}\approx x_{ts}-f_S/F''(x_{ts})$. 
For a sharp transition barrier ($x_{ts}F''(x_{ts})\gg f_S$), 
the force will not affect the position of the transition state ($x_{ts}^{NEW}\approx x_{ts}$).
If the transition barrier is broadly distributed as in 
the unzipping pathway of RNA hairpins, the structure of transition state progressively changes 
as the magnitude of the force is varied.
Typically, folding transition states are shallow and broad. 
As a result in biomolecular folding or unbinding, which involve formation or rupture of non-covalent 
interactions, we predict that the location of the TS depends on $f_S$ and temperature. 
The assumption of a fixed TS used to interpret \cite{Bustamante2,AjdariBJ04,Evans1} experimental results is not valid. 
In addition, sequential \cite{EvansNature99} and/or 
parallel pathways to the stretched state transition \cite{NevoNSB03,BarsegovPNAS05} are also possible. 
These observations suggest that 
a careful inspection of nonlinearity in the Arrhenius plot and $\alpha_f$ will 
be required to unravel barriers to unfolding.\\

{\bf Entropic barriers and long refolding times from fully stretched state: }
We have proposed that the long refolding time in P5GA upon force-quench from the initial stretched conformations 
is due to entropic barriers.
The rate limiting step in the force-quench refolding of P5GA is 
the $t\rightarrow g^+$ transition in the nucleotides near the GAAA tetraloops.
We analyze the simulation results by adopting a model proposed by Zwanzig \cite{ZwanzigPNAS95}. 
In this Ising-like model each degree of freedom (in our case the dihedral angle) is presumed to 
exist in the ``correct'' state (native) and ``incorrect'' or non-native state. 
Suppose that the energy difference between the incorrect and correct states 
is $\epsilon$ and that 
there are $N$ correct dihedral angles required for forming the hairpin loop. 
Let the free energy of loop stabilization be $\epsilon_{loop}$. 
The energy, $E_n$, of a conformation with $n$-incorrect dihedral angles is
$E_n=n\epsilon-\delta_{n0}\epsilon_{loop}$.
If we assume that the dihedral angle is a discrete variable with $1+\nu$ states ($\nu$ is the number of incorrect states) then   
the partition function is 
\begin{equation}
Z_N=\sum_{n=0}^N{N\choose n}\nu^ne^{-\beta(n\epsilon-\delta_{n0}\epsilon_{loop})}=e^{\beta\epsilon_{loop}}+(1+\nu e^{-\beta\epsilon})^N-1.
\end{equation}
The thermal probability of realizing a conformation with $n$ incorrect dihedral angles is 
\begin{equation}
P_n(eq)=\frac{{N\choose n}\nu^ne^{-\beta(n\epsilon-\delta_{n0}\epsilon_{loop})}}{Z_N}. 
\end{equation}
The free energy profile, with $n$ playing the role of a reaction coordinate for dihedral angle transitions, 
is $\Delta F(n)=n\epsilon-\delta_{n0}\epsilon_{loop}-k_BT\log{\nu^n{N\choose n}}$.
For P5GA $N=6$, $\epsilon\approx 2k_BT$, and $\epsilon_{loop}\approx 7.6k_BT$ ($=V_{STACK}^{B_8B_9B_{14}B_{13}}+V_{LJ}^{B_9B_{14}}=5.1k_BT+2.5k_BT$). 
The free energy profile  $\Delta F(n)$ (Fig.\ref{Zwanzig}) shows that the barrier depends only weakly on $\nu$. 
Because the dihedral angle is a continuous variable, we use $\nu$ as an undetermined parameter.  
For $\nu=13$ the free energy barrier ($\Delta F_F^{\ddagger}$) is $\sim 2.7 k_BT$, which leads to 
the observed increase ($\tau_F(f_Q=0)=15\tau_F(T)$) in the refolding time by factor of $15$ ($=e^{\Delta F_F^{\ddagger}/k_BT}$).

The entropic barrier $\sim (5-6) k_BT$ is significantly larger than the 
free energy barrier $\Delta F^{\ddagger}_U$ in the absence of force. 
The barrier to the formation of conformations with ($tttttt$) state in the loop region is large enough that they do not form
by thermal fluctuations, and hence are irrelevant when refolding is initiated by temperature quench. 
However, such conformations are populated with near unit probability when fully stretched by 
mechanical force. 
When folding is initiated by force-quench from extended conformations (like conformation I in Fig.\ref{dihedral}-E), 
metastable conformations with incorrect dihedral angles in the loop regions are formed. 
These are characterized by plateaus in the dynamics of $R$ (Fig.\ref{evolutionDIH}-A). 
The crossing of the entropic barrier that places the loop dihedral angles in the native-like 
gauche state results in the slow refolding of P5GA hairpins. 
Once the loop is formed the zipping process quickly stabilizes the hairpin so that barrier crossing in reverse direction is unlikely to occur at low forces.

To further validate the proposed mechanism we performed simulations in which the value of the initial stretching force $f_S$ is not large enough 
to fully extend the P5GA hairpin. 
In these simulations, the ensemble of initial structures is prepared so that they contain the preformed GAAA loop 
with only single bond 
before the loop region that is intact (see conformation II in Fig.\ref{dihedral}-E). 
The refolding kinetics follows a 
single exponential decay with a mean refolding time of $\sim 33$ $\mu s$, 
which is only 2-3 times longer than the refolding time of thermally denatured states. 
This value is much shorter than the refolding time from the fully stretched states ($>191$ $\mu s$, see Fig.\ref{unfoldrefold}-B).
These simulations also show that $\tau_F$ is a function of both $f_S$ and $f_Q$.

A byproduct of this analysis is that the appropriate reaction coordinate in force-quench refolding of RNA hairpins may be a local variable.
In the formation of P5GA hairpin the local dihedral angles are the relevant reaction coordinates. 
The local dihedral coordinates, that describe the rate-limiting steps in 
the UBA$\rightarrow$NBA transition, are \emph{hidden} in the global coordinates such as $Q$ or $R$. 
Indeed, there is no correlation between the formation of native dihedral angles in the GAAA tetraloop and global order parameters. 
We infer that to describe folding, especially of RNA, multiple reaction coordinates that describe the hierarchical assembly are required. \\

{\bf Difficulties in extracting energy landscape parameters from single molecule force spectroscopy: }
Several studies \cite{HummerBP03,AjdariBJ04} have pointed out the inherent ambiguities in quantitatively characterizing the energy 
landscape from measurable quantities in dynamic force spectroscopy. 
From the plots [$f^*$,$\log{r_f}$] one cannot unambiguously obtain the location 
of the transition state(s) or even the number of free energy barriers \cite{AjdariBJ04}. 
The significant curvatures in the [$f^*$,$\log{r_f}$] plots are usually interpreted in terms of multiple transition states \cite{EvansNature99,AjdariBJ04}. 
In our example, the P5GA hairpin unfolds upon application of force by \emph{crossing a single free energy barrier}. 
Explicit equilibrium $F(R)$ profiles (Fig.\ref{unfoldrefold}-C) and experiments \cite{Bustamante2} 
that have monitored hopping dynamics 
in P5ab hairpin show that there is only one free energy barrier in these simple structures. 
Nevertheless, [$f^*$,$\log{r_f}$] plot is highly nonlinear (Fig.\ref{pullsummary}-B). 
In the hairpin case we have shown that the non-linearity is due to the dramatic changes in $\Delta x^{TS}_F$ as $r_f$ is varied. 
The standard assumption that $\Delta x_U^{TS}$ is a constant breaks down, 
and is likely to be an even more of a severe approximation 
for RNA with tertiary structures. 

To further illustrate the importance of transition state movements we consider a trivial one dimensional potential 
\begin{equation}
E(x)=-\epsilon\exp{(-\xi x)}.
\end{equation}
In this barrierless potential a particle is ``unbound'' if $|E(x_{ts})/k_BT|<1$ where $x_{ts}$ is the TS location. 
Upon application of a constant $f_S$ the potential becomes 
\begin{equation}
E(x)=-\epsilon\exp{(-\xi x)}-f_Sx.
\end{equation}
The location of the transition state in the force range $0<f_S<\xi\epsilon$ is 
\begin{equation}
x_U^{TS}=-\frac{1}{\xi}\log{f_S/\xi\epsilon}.
\end{equation}
If $f_S>\xi\epsilon$ then $x^{TS}_U(f)=0$ and the particle is always unbound (Fig.\ref{exp_TS_shift}-A).
The changes in $x_U^{TS}$ can lead to significant deviations from the Bell equation even in constant $f_S$-experiments. 
The distribution of unbinding forces upon deforming the potential at constant loading rate ($f_S(t)=r_ft$) can be 
analytically obtained (see Eq.(8) in \cite{HyeonPNAS03}). 
The [$f^*$,$\log{r_f}$] plot from these calculations show non-linearities (Fig.\ref{exp_TS_shift}-B) that are similar to 
what is found for P5GA (Fig.\ref{pullsummary}-B). 
In both instances the reason for curvature is entirely due to $r_f$-dependent changes in the TS location. 
\\

{\bf CONCLUSIONS}\\

We have systematically investigated forced-unfolding and force-quench refolding of RNA hairpins. 
Using a general minimal model for RNA 
we have obtained a number of new results 
that give a molecular picture of unfolding and refolding of RNA hairpins triggered by force. 
Although they were obtained specifically for P5GA we expect the conclusions to be valid for other RNA sequences as well. 
The specific predictions, that are amenable to experimental test, of our study are listed below. 
\begin{description}
\item[1)] Besides probing the energy landscape using $f$ as a ``denaturant'' one 
of the goals of single molecule studies is to extract intrinsic parameters like folding ($k_F^o$)
and unfolding rates ($k_U^o$) and the nature of the transition state in the absence of 
force. 
However, extraction of the kinetic parameters from dynamic force spectroscopy is fraught with difficulties 
because several models can produce similar [$f^*$,$\log{r_f}$] profiles. 
Here we have shown, for a system \emph{that has only a single barrier to unfolding}, 
that $\Delta x^{TS}_U$ depends dramatically on $f_S$ and $r_f$. 
The movements in the TS is intrinsic to properties of RNA hairpins
and are clearly reflected in the free energy profiles. 
Combining the present results and our previous study \cite{HyeonPNAS05} 
we surmise that $\Delta x^{TS}_U$ is dependent on $T$ and $f$. 
Thus, extrapolations to zero force to obtain reliable estimates of unfolding rates requires not only accounting for 
free energy barriers \cite{HummerBP03} but also on the dependence of $\Delta x^{TS}_U$ on $T$ and $f$. 
Only by performing multiple experiments (or simulations) over a range of $f_S$ and $T$ can 
the free energy landscape be fully characterized. 
\item[2)] An important prediction of our work is that refolding times upon force-quench $\tau_F(f_Q)$ from stretched states are much 
greater than those obtained by temperature quench. 
More generally, $\tau_F(f_Q)$ depends, sensitively on the initial value of the 
stretching force. The microscopic origin of the long force-quench refolding times in P5GA has been traced to the 
time needed for the trans$\rightarrow$gauche transition in the GAAA tetraloop region. 
From this observation we predict that refolding time $\tau_F(f_Q)$ should be very long compared to thermal refolding times for P5ab which also has the GAAA tetraloop. 
Because the refolding times for RNA hairpins are determined by the local structural features in the initial stretched 
states we suggest that, at a fixed temperature, $\tau_F$($f_Q$) might \emph{depend upon only weakly} 
on the helix length or the precise sequence (percentage of GC for example). 
\item[3)] Dissecting the folding mechanism of RNA is difficult because of an interplay of a number 
of factors \cite{HyeonBC05}. 
We predict that refolding mechanisms (pathways and the nature of the transition state ensemble) by temperature (or by increasing cation concentration) and force quench have to be drastically different. 
In the former, the transition to the low entropic NBA proceeds from a high entropy relatively compact state whereas in the latter it occurs from a low entropy stretched state \cite{LiPNAS06}. 
The predicted dramatic differences in the folding mechanisms can be established by probing force-quench refolding at fixed $T$ and counterion concentration. 
\end{description}

The present model has a number of limitations.  
The use of Go model for force-unfolding may not be a serious approximation 
because unfolding pathway are largely determined by the native topology \cite{Klimov2}. 
However, the neglect of non-native interactions will have dramatic effect on refolding. 
At a minimum, the roughness ($\delta\epsilon$) of the energy landscape is underestimated by the Go model. 
The importance of $\delta\epsilon$ can be assessed by doing forced-unfolding experiments over a range of temperature \cite{HyeonPNAS03,ReichEMBOrep05}. 
Finally, the electrostatic interactions in RNA have been modeled in the simplest manner that is only appropriate for 
monovalent cation \cite{LeeEJB99}. 
To address the effect of counterion (Mg$^{2+}$ or polyanions) explicit modeling of the cations will be required. \\

{\bf APPENDIX I}\\

In this appendix we describe the procedure for determining the persistence length of the linkers used in our simulations. 
For the linker molecules, whose energy function is given by  
Eq.(\ref{eqn:Linker_Hamiltonian}), we calculated the persistence length by fitting 
the worm-like chain end-to-end ($R$) distribution function ($P_{WLC}(R)$) \cite{HaBook} 
to the simulated $P(R)$. 
We adopted Monte Carlo simulation with Pivot algorithm \cite{BishopJCP91} 
to generate a large number of equilibrium conformations of the linker molecule. 
Given $k_B(=20$ $kcal/(mol\cdot\AA^2))$, $k_A$($=20$ $kcal/mol$ or $80$ $kcal/mol$), and varying $N$ the number of monomers in the WLC linker, 
we obtained the unknown parameters, namely, the contour length ($L$) and the persistence length ($l_p$) by fitting $P(R)$ to 
\begin{equation}
P_{WLC}(R) = \frac{4\pi C(R/L)^2}{L[1-(R/L)^2]^{9/2}}\exp[-\frac{3t}{4(1-(R/L)^2)}].
\label{eqn:WLC}
\end{equation}
where $t\equiv L/l_p$.
The normalization constant
$C=1/[\pi^{3/2}e^{-\alpha}\alpha^{-3/2}(1+3\alpha^{-1}+15/4\alpha^{-2})]$ with $\alpha=3t/4$, 
satisfies $\int_0^LdRP_{WLC}(R)=1$. 
The dependence of the persistence length of the linkers, as a function of $N$ is displayed in Fig.\ref{lp_comp}.  
The quality of the fit improves as $N$ becomes larger (data not shown). 
We also computed the persistence length and the contour length 
of P5GA at $T>T_m$ using the same fitting procedure, which gives $l_p^{RNA}\approx 1.5$ $nm$ and $L=12.5$ $nm$ (Fig.\ref{lp_comp}-inset on the bottom). 
In our simulations $\lambda=l_p^H/l_p^{RNA}$ ranges from $10<\lambda<70$. 
The experimental value of $\lambda\approx 50$ \cite{Bustamante2}\\ 

{\bf Acknowledgments:} This work was supported in part by a grant from the National Science Foundation (CHE 05-14056).\\

{\bf Note added:} 
While the present paper was under review refolding upon 
force-quench of TAR RNA was reported \cite{TinocoBJ06}. 
In accord with the present and our previous studies \cite{HyeonPNAS05},  
force-quench refolding times are relatively long. 
The distributions of refolding times similar to the curves in 
Fig.\ref{lagtimeanalysis}, as a function of $f_Q$ were not reported in 
\cite{TinocoBJ06}. 
Thus, it is unclear if there is a lag phase in the force quench refolding of TAR RNA.
\newpage

\newpage

\section*{\bf FIGURE CAPTION}

\begin{figure}[ht]
\includegraphics[width=5.50in]{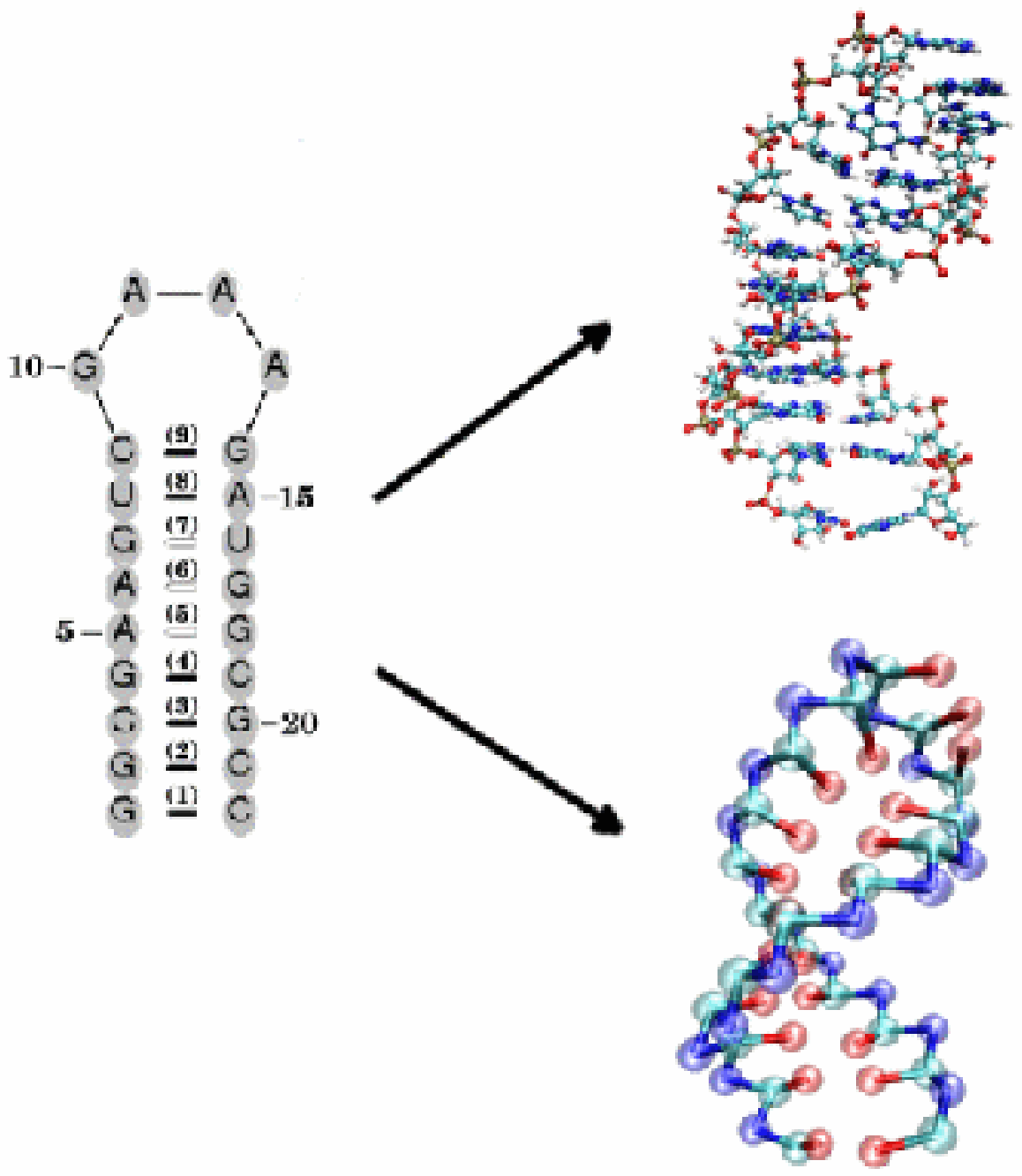}
\caption{\label{coarse_RNA}}
\end{figure}

{{\bf Figure} \ref{coarse_RNA} :}
Coarse-grained representation of a RNA using three (phosphate (P), sugar (S) and base (B)) 
interaction sites per nucleotide. 
On the left we present the secondary structure of the 22-nt P5GA hairpin in which the bonds formed
between base pairs are labeled from 1 to 9.
The PDB structure \cite{TinocoJMB2000} and the lowest energy structure obtained with the 
coarse-grained model are shown on the right.

\begin{figure}[ht]
\includegraphics[width=5.00in]{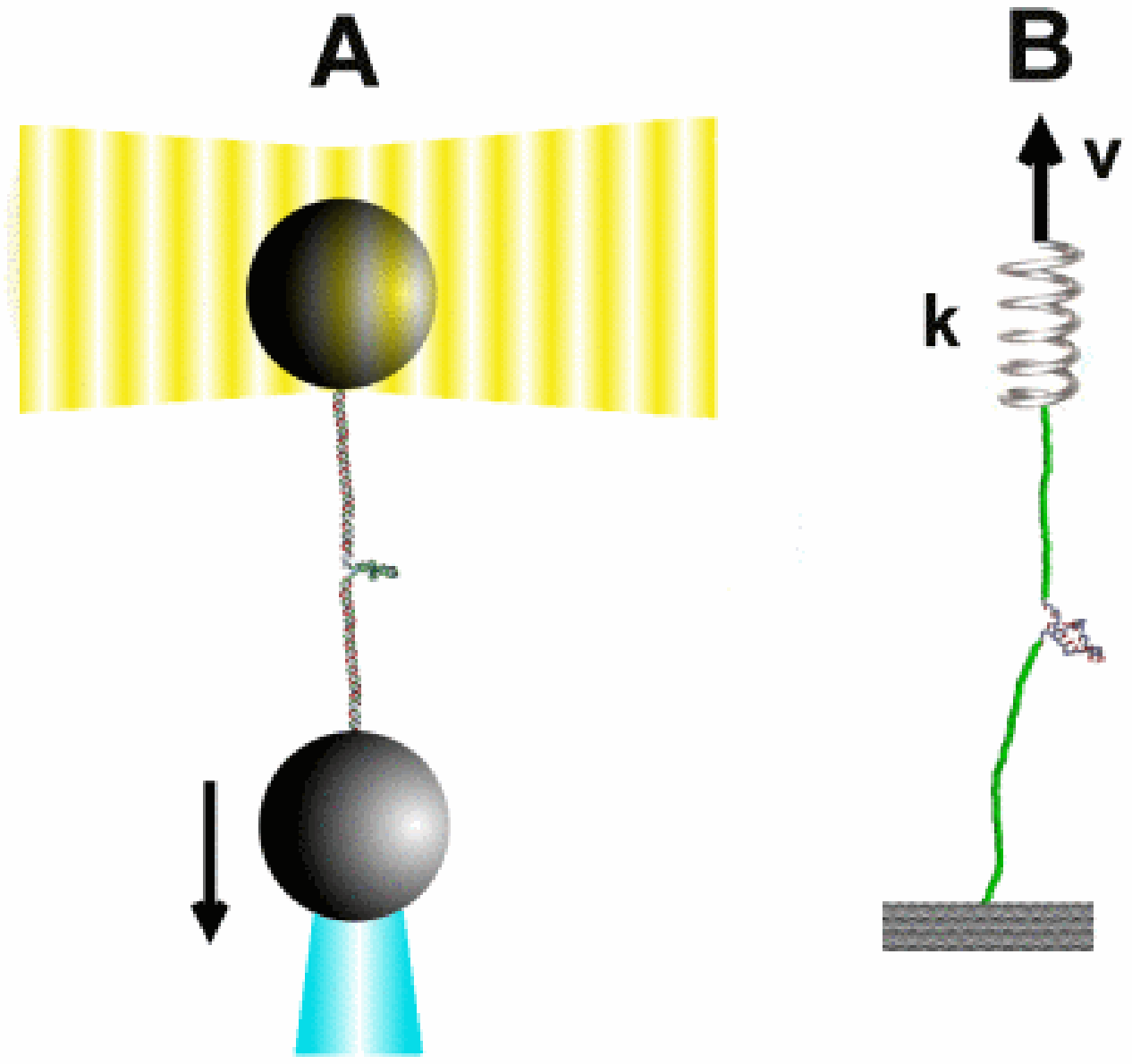}
\caption{\label{exp}}
\end{figure}

{{\bf Figure} \ref{exp} :}
{\bf A.} Schematic illustration of laser optical tweezer (LOT) setup for RNA stretching. 
Single RNA molecule is held between two polystyrene beads via 
molecular handles with one of the polystyrene beads being optically trapped in the laser light. 
The location of the other bead is changed by manipulating it by a micropipette. 
The extension of the molecule through the molecular handles 
induces the deviation in the position of the polystyrene bead held in the force-measuring 
optical trap.
{\bf B.} Both LOT and AFM can be conceptualized as schematically shown. 
The RNA molecule is sandwiched between the linkers and one end of the linker is pulled. 
The spring constant of the harmonic trap in LOT (or the cantilever in AFM experiments) is 
given by $k$ and $v$ is the pulling speed. 

\begin{figure}[ht]
\includegraphics[width=5.0in]{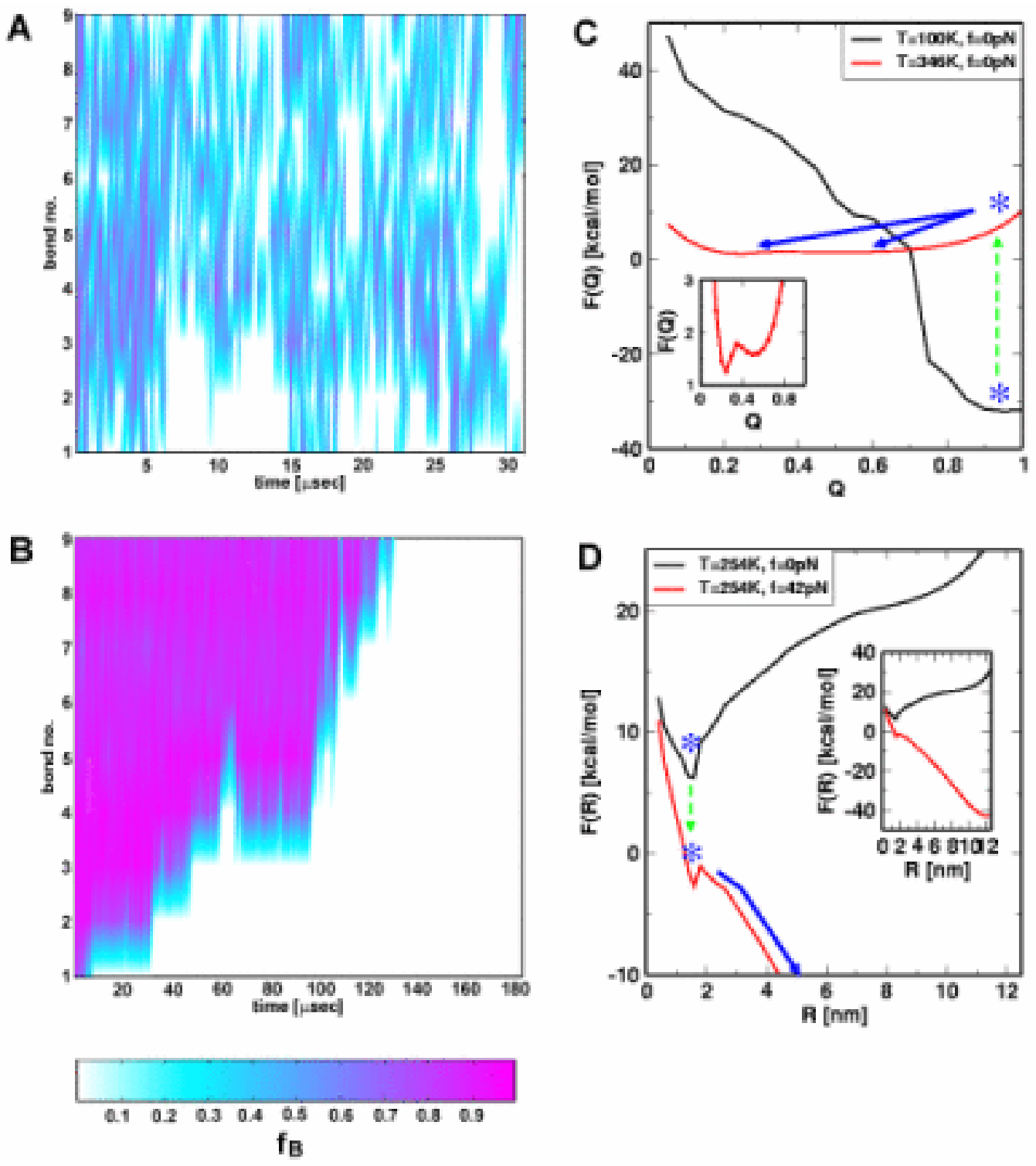}
\caption{\label{unfoldpath}}
\end{figure}

{{\bf Figure} \ref{unfoldpath} :}
Unfolding pathways upon temperature and force jump.
{\bf A.} The time dependence of rupture of the bonds is monitored when the temperature is raised from $T(=100$ $K)<T_m(\approx 341$ $K)$ to $T(=346$ $K)>T_m$.
The set of nine bonds are disrupted stochastically. 
{\bf B.} In forced unfolding bonds rip from the ends to the loop regions in an apparent staircase pattern. For both {\bf A} and {\bf B} the scale indicating the probability of a given bond being intact is given below. 
{\bf C.} Free energy $F(Q)$ as a function of $Q$. 
The stable hairpin with $Q\approx 1$ at $T=100$ $K$ becomes unstable upon rapid temperature jump to $T=346$ $K$ (blue *). 
Subsequent to the T-jump the hairpin relaxes to the new equilibrium state by crossing a small free energy barrier ($\approx 0.5$ $kcal/mol$) with $Q\approx 0.2$. 
The inset shows the equilibrium free energy profile at $T=346$ $K$ and $f=0$. 
{\bf D.} Deformation of the free energy profile upon application of force. The $R$ dependent 
free energy $F(R)=F(R;f_S=0)-f_S\cdot R$ favors the stretched state at $R=12$ $nm$ when $f=42$ $pN$ (see inset). 
The activation barrier separating the UBA and NBA is around (1-2)$kcal/mol$.

\begin{figure}[ht]
\includegraphics[width=6.00in]{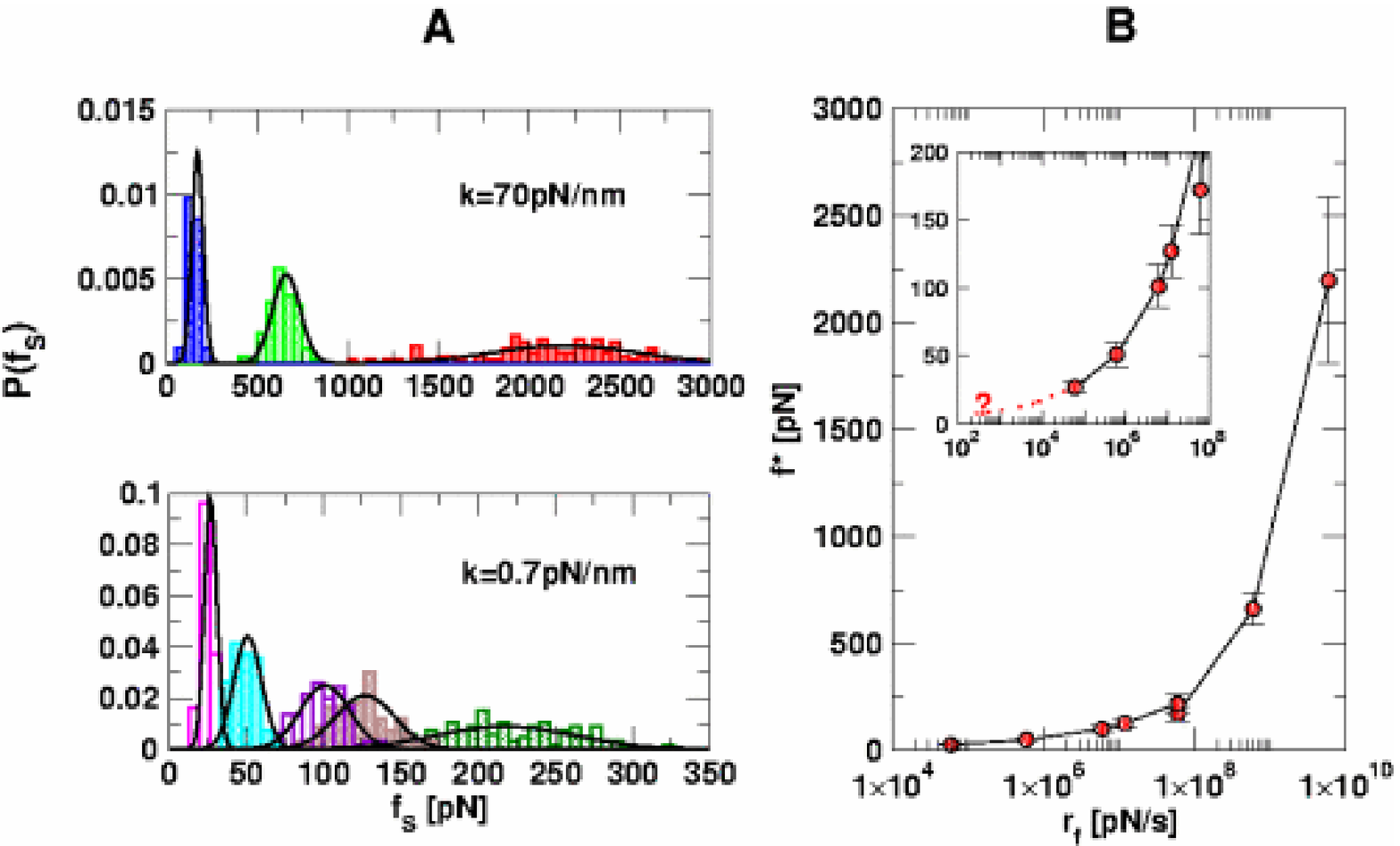}
\caption{\label{pullsummary}}
\end{figure}

{{\bf Figure} \ref{pullsummary} :}
Constant loading rate force unfolding.
{\bf A.} The unfolding force distributions at different pulling speeds with hard ($k=70$ $pN/nm$, up) 
and soft springs ($k=0.7$ $pN/nm$, down). 
For the hard spring, the pulling speeds from right to left are 
$v=8.6\times 10^4$, $8.6\times 10^3$, $8.6\times 10^2$ $\mu m/s$.
For the soft spring, the pulling speeds are 
$v=8.6\times 10^4$, $1.7\times 10^4$, $8.6\times 10^3$, $8.6\times 10^2$, $8.6\times 10^1$ $\mu m/s$ from right to left force peaks. The peak in the distributions which are fit to a Gaussian is the most probable force 
$f^*$.
{\bf B.} The dependence of $f^*$ as a function of the 
loading rates, $r_f$. 
The results from the hard spring and soft spring are combined using the loading rate as the relevant variable. 
The inset illustrates the potential difficulties in extrapolating from simulations at large $r_f$ to small values of $r_f$. 

\begin{figure}[ht]
\includegraphics[width=6.00in]{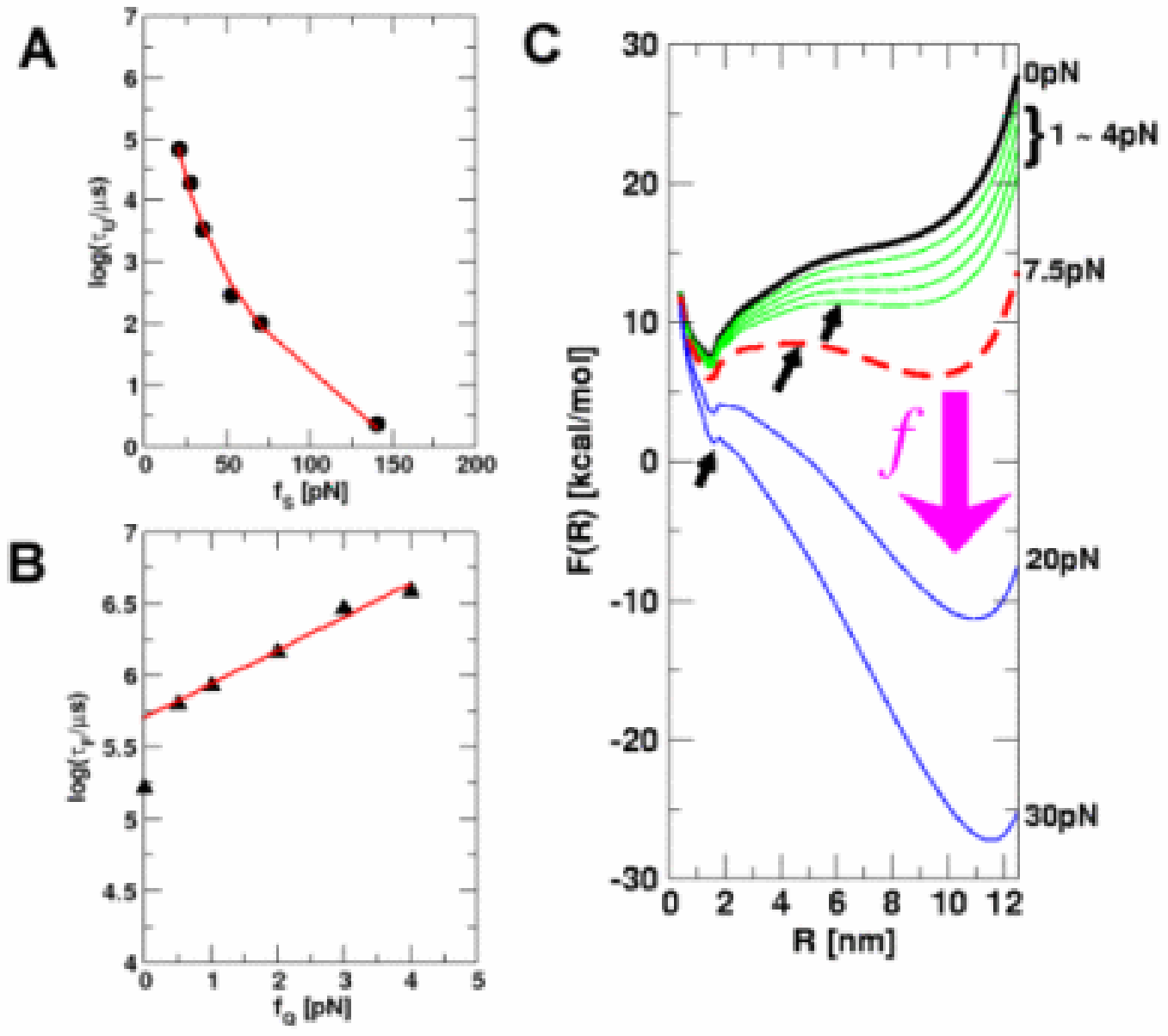}
\caption{\label{unfoldrefold}}
\end{figure}

{{\bf Figure} \ref{unfoldrefold} :}
Kinetics of forced-unfolding and force-quench refolding: {\bf A.} Plot of force-induced unfolding times ($\tau_U$) as a function of the stretching force. 
Over a narrow range of force $\tau_U$ decreases exponentially as $f$ increases. 
{\bf B.} Refolding time $\tau_F$ as a function of $f_Q$. 
The initial value of the stretching force is 90 $pN$. By fitting $\tau_F$ using $\tau_F(f_Q)=\tau_F^o\exp{(f_Q\Delta x_F^{TS}/k_BT)}$, in the range of $0.5$ $pN<f_Q<4$ $pN$,
we obtain $\Delta x_F^{TS}\approx 1$ $nm$ and $\tau^o_F\approx 290$ $\mu s$. 
{\bf C.} Changes in the equilibrium free energy profiles at $T=290$ $K$ $F(R)$ as a function of the variable $R$. 
We show $F(R)$ at various $f_S$ values. 
For emphasis, the free energies at $f_S=0$ and at the transition midpoint $f_S=7.5$ $pN$ (dashed line) are drawn in thick lines.

{{\bf Figure} \ref{lagtimeanalysis} :}
Time dependence of the probability that RNA is unfolded upon force quench. 
In these simulation $T=290$ $K$, and the initial stretching force $f_S=90$ $pN$ and $f_Q$ (values are given in each panel), the quench force, is varied. The simulation results are fit using Eq.(\ref{eqn:consecutive_process}) which is obtained using the kinetic scheme $S \stackrel{\tau_1}\longrightarrow I\stackrel{\tau_2}\rightarrow F$. 
Here, $I$ represents conformations with in certain fraction of incorrect dihedral angles.
The time constants ($\tau_1$,$\tau_2$), in $\mu s$, at each force are: 
(81.2, 101.3) at $f_Q=0$ $pN$, 
(159.5, 160.8) at $f_Q=0.5$ $pN$, 
(180.0, 174.8) at $f_Q=1$ $pN$, 
(237.6, 240.5) at $f_Q=2$ $pN$, 
(326.8, 335.6) at $f_Q=3$ $pN$, and 
(347.7, 329.7) at $f_Q=4$ $pN$. 

{{\bf Figure} \ref{Ext_refold} :}
{\bf A.} The equilibrium distribution of the end-to-end distance at extremely high temperature ($T=1500$ $K$). 
Even at this elevated temperature the fully stretched conformations of $R=13.5$ $nm$ (arrow) 
is not found in the ensemble of thermally denatured conformations.
{\bf B.} Refolding is initiated by a force quench from the initial value $f_S=90$ $pN$
to $f_Q=4$ $pN$.
The five time traces show great variations in the relaxation to the hairpin conformation.
However, in all trajectories $R$ decreases in at least three distinct stages that are explicitly labeled for the trajectory in green.

{{\bf Figure} \ref{dihedral} :}
{\bf A.} The dihedral angles of the P5GA hairpin in the native state. All the dihedral angles are in the 
trans form except 19-th position of dihedral angle which is in the gauche(+) conformation (indicated by orange circle).
{\bf B.} The dihedral angle potentials for trans (top) and gauche(+) form (bottom) are plotted using Eq.(\ref{eqn:DIH}). The red lines show the potentials in the loop region. 
{\bf C, D.} The average deviation of the $i$-th dihedral angle relative to 
the native state is computed using the 100 different structures generated by high temperature ($T=500$ $K$) ({\bf C}) 
and by force ($R=13.5$ $nm$) ({\bf D}).
To express the deviations of the dihedral angles from their native state values we used $1-\cos{(\phi_i-\phi_i^o)}$ for 
$i$-th dihedral angle $\phi_i$ where $\phi_i^o$ is the $i$-th dihedral angle of the native state.
{\bf E.} A snapshot of fully stretched hairpin (I). Note the transition in 19-th dihedral angle undergoes $g^+\rightarrow t$ transition when hairpin is stretched. An example of a partially stretched conformation (II) with the GAAA tetraloop and bond 9 intact.
Refolding times starting from these conformations are expected to be shorter than those that start from fully stretched states.

{{\bf Figure} \ref{evolutionDIH} :}
{\bf A.} A sample refolding trajectory starting from the stretched state. 
The hairpin was initially unfolded to a fully stretched state 
and $f_Q$ was set to zero at $t\approx 20$ $\mu s$. 
End-to-end distance monitored as a function of time shows that refolding occurs in steps. 
{\bf B.} The deviation of dihedral angles from their values in native state as a function of  time. The large deviation of 
the dihedral angles in loop region can be seen in the red strip. 
Note that this strip disappears around $t\approx 300$ $\mu s$, which coincides with the formation of bonds 
shown in {\bf C}. $f_B$ is the fraction of bonds with pink color indicating that the bond is fully formed.

{{\bf Figure} \ref{linker_anal} :}
The force extension curves (FECs) of RNA hairpin at constant pulling speed and 
varying linker lengths and flexibilities.
The pulling speed, $v=0.86\times 10^2$ $\mu m/s$. 
The spring constant, $k=0.7$ $pN/nm$.
FECs from different linker lengths are plotted in black (10 $nm$), red (20 $nm$), green (30 $nm$),
blue (40 $nm$), and orange (50 $nm$).
{\bf A.} This panel shows the experimentally relevant plots, namely, FECs for H-RNA-H construct. 
The signature for the hairpin opening transition region is ambiguous at large $L_H$ values. 
{\bf B.} FECs only for P5GA corresponding to different $L_H$ values. The 
FEC for the linker is subtracted from (A). 
The gradual increase of rupture force is observed as L increases.
The value of $k_A$ (Eq.(\ref{eqn:Linker_Hamiltonian})) of linker polymer used in ({\bf A}) and ({\bf B}) is 80 $kcal/(mol\cdot\AA)$.
{\bf C.} Comparison between FECs with $L_H(=40$ $nm)$ fixed but at different $k_A$ values. 
Red curve is the average of 7 individual FECs that are shown in orange (80 $kcal/(mol\cdot\AA)$). 
Black curve is the average of 9 individual FECs in grey (20 $kcal/(mol\cdot\AA)$).
Less stiff linker leads to slightly larger unfolding force. It should be stressed that for both values of $k_A$ the $\lambda$ ratio is large. 

{{\bf Figure} \ref{refoldlinker} :}
Refolding trajectory of RNA that is attached to the handles. 
Linkers with $L_H=15$ $nm$ and $l_p^H=30$ $nm$ are attached to both side of RNA hairpin.  
The initial force ($90$ $pN$) stretched P5GA to 14 $nm$. 
The value of $f_Q=2$ $pN$. The top panel shows the dynamics of the end-to-end distance, $R_{sys}$, of H-P5GA-H. 
The middle panel corresponds to the end-to-end distance of P5GA. The end-to-end distances, $R_H$, of the 3' and 5'-side handles fluctuates around its equilibrium value of $14$ $nm$ after the initial rapid relaxation. 
The right panels show the decomposition of $R_H$ into the longitudinal and the transverse 
components.
The value of $R^{||}_H\approx 13$ $nm$ agrees well with the equilibrium value obtained by solving 
$f_Q=k_BT/l_p^H\left(R_H^{||}/L_H+1/4(1-\frac{R_H^{||}}{L_H})^2-1/4\right)$ with $f_Q=2pN$.

{{\bf Figure} \ref{Hammond} :}
{\bf A.} Free energy profiles of $Q$ at different temperatures. Note that the positions of the transition states over the temperature variation almost remain constant.
{\bf B.} The movement of transition state measured in terms of the Leffler parameter (Eq.(\ref{eqn:Lefler})). 
The structural nature of transition state is 
monitored by the free energy difference between NBA and UBA when $f$ is a external variable.
The inset shows the variation of $\alpha$ with respect to force. The value of $T=290$ $K$.

{{\bf Figure} \ref{Zwanzig} :}
The free energy as a function of the number of incorrect dihedral angles calculated using the Zwanzig model (see text for details). 
As the number of distinct values ($\nu$) that the dihedral angle can take increases the 
entropic barrier increases. 
For P5GA $\nu=13$ provides the best fit of the model to simulations (see text). 

{{\bf Figure} \ref{exp_TS_shift}:}
{\bf A.} Sketch of the one-dimensional potential $E(x)$ as a function of $x$ for several values of $f_S$. 
The transition state location is obtained using $E'(x_{ts})=0$. 
The boundary separating bound and unbound states is given by $|E(x_{ts})/k_BT|=1$.
{\bf B.} Dependence of the most probable ``unbinding'' force $f^*$ as a function of $r_f$. 
The [$f^*$,$\log{r_f}$] plot for the artificial potential is similar to that shown in Fig.\ref{pullsummary}-B.

{{\bf Figure} \ref{lp_comp} :}
Persistence length $l_p$ as a function of contour length $L$ for linkers at $290$ $K$. 
The number of monomers $N$ (Eq.(\ref{eqn:Linker_Hamiltonian})) is also shown. 
The values of $l_p$ and $L$ are obtained by fitting the end-to-end distance distribution function 
$P(R)$ generated by simulations to the theoretical expression based on a mean field model (Eq.(\ref{eqn:WLC})). 
An example of such a fit for a linker with $N=50$ 
using two different values of $k_A$ (see Eq.(\ref{eqn:Linker_Hamiltonian}))
is shown in the inset on the top. 
For large $N$, $l_p$ converges to a constant value. 
The fit of simulated $P(R)$ for P5GA computed at $T(=500$ $K)>T_m$ is shown in the inset on the bottom. From the WLC fit we obtain $l_p\approx 1.5$ $nm$ and $L=12.5$ $nm$ for P5GA. 

\newpage
\newpage
\newpage
\newpage 
\newpage
\newpage 
\begin{figure}[ht]
\includegraphics[width=6.00in]{lagtimeanalysis.eps}
\caption{\label{lagtimeanalysis}}
\end{figure}
\newpage
\begin{figure}[ht]
\includegraphics[width=4.0in]{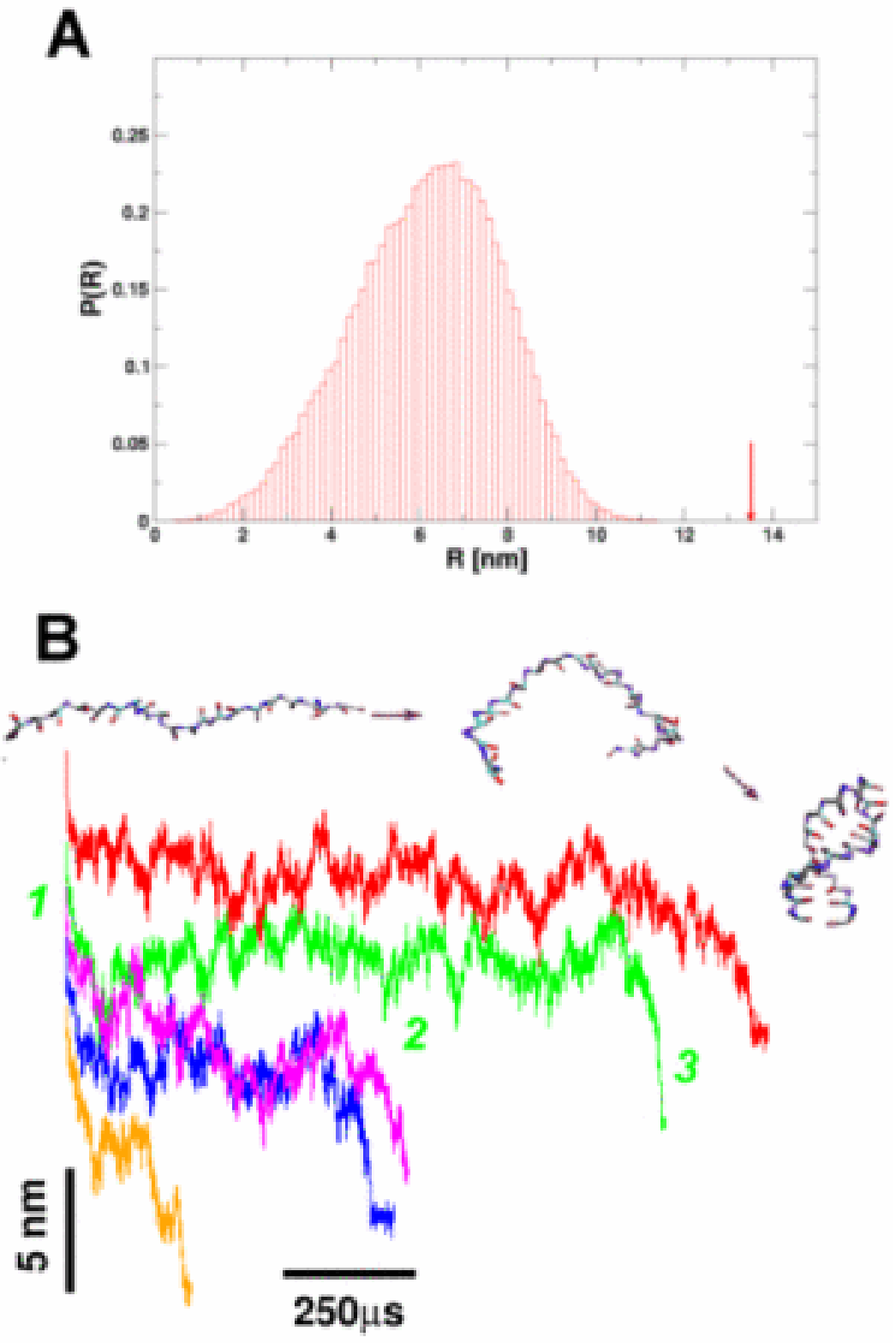}
\caption{\label{Ext_refold}}
\end{figure}
\begin{figure}[ht]
\includegraphics[width=5.50in]{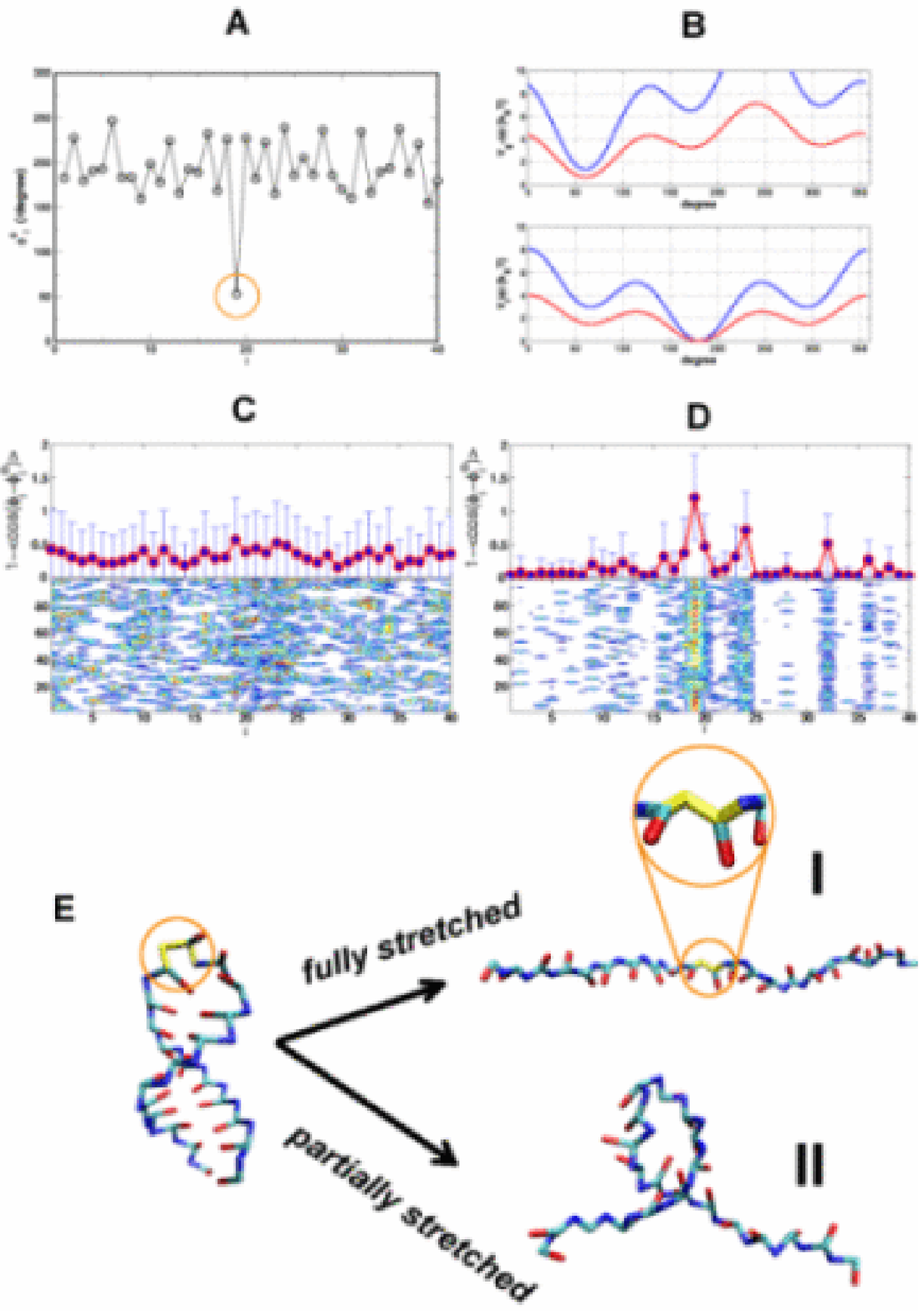}
\caption{\label{dihedral}}
\end{figure}
\newpage
\begin{figure}[ht]
\includegraphics[width=7.00in]{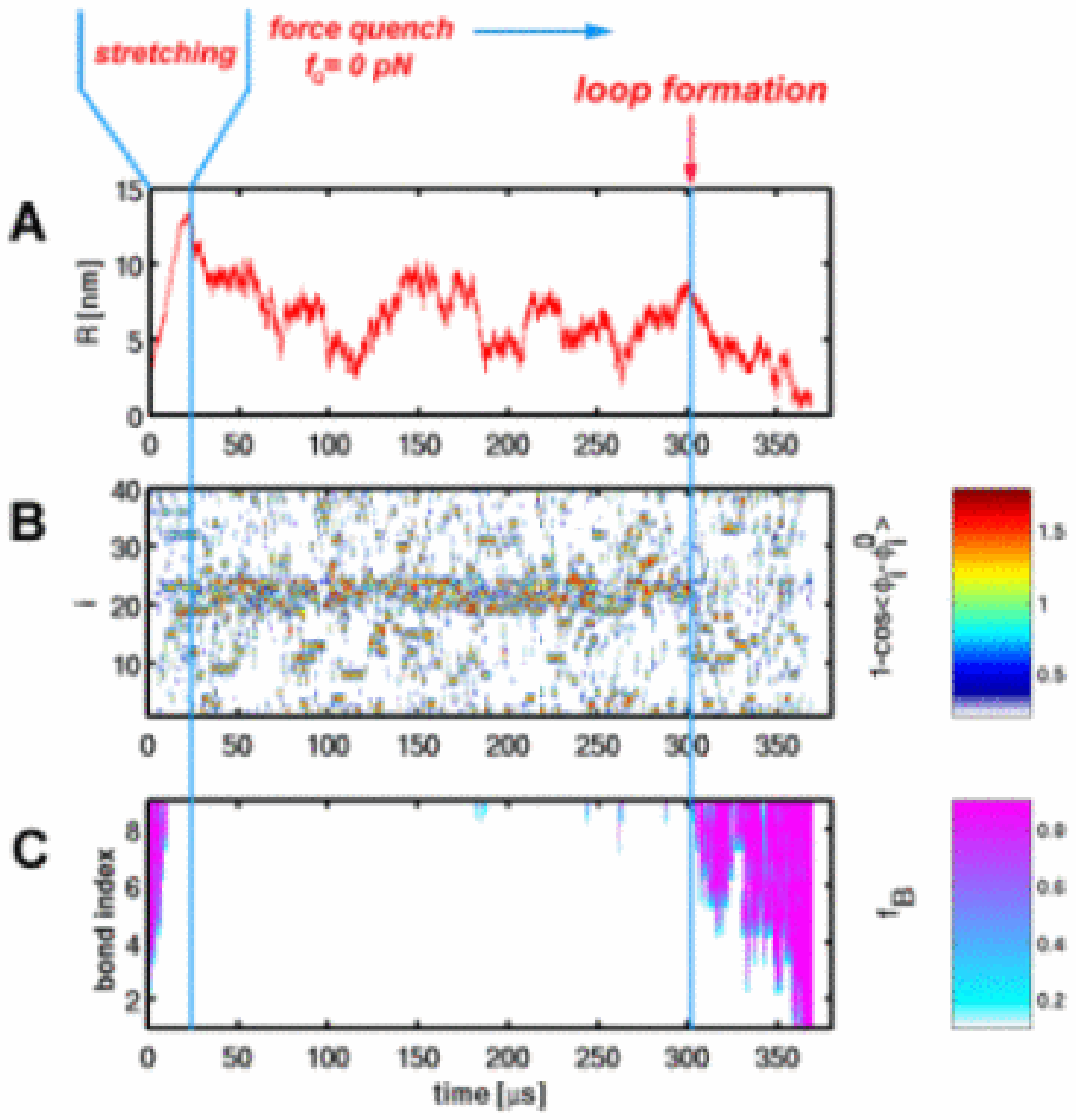}
\caption{\label{evolutionDIH}}
\end{figure}
\newpage
\begin{figure}[ht]
\includegraphics[width=6.00in]{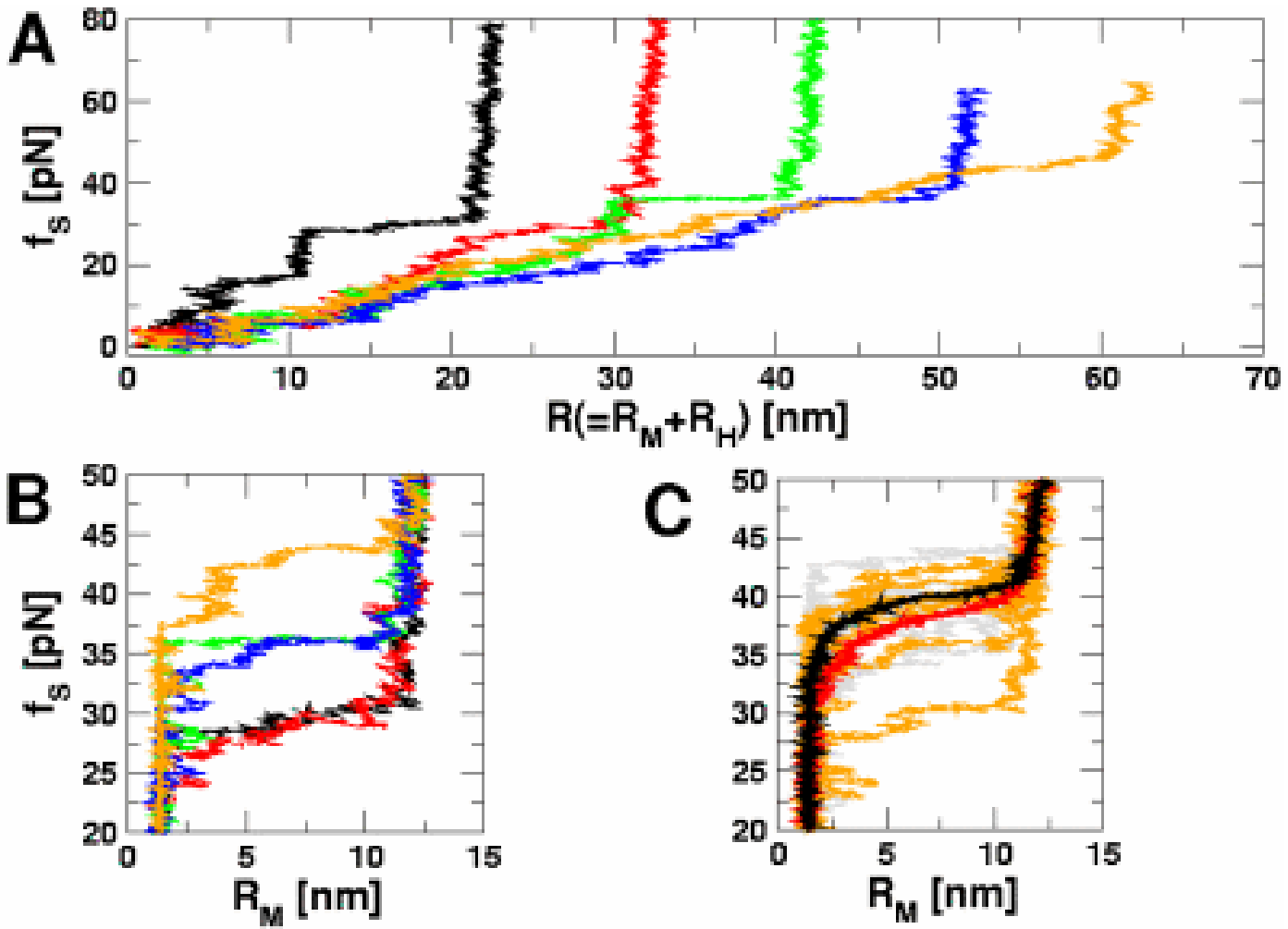}
\caption{\label{linker_anal}}
\end{figure}
\newpage
\begin{figure}[ht]
\includegraphics[width=5.00in]{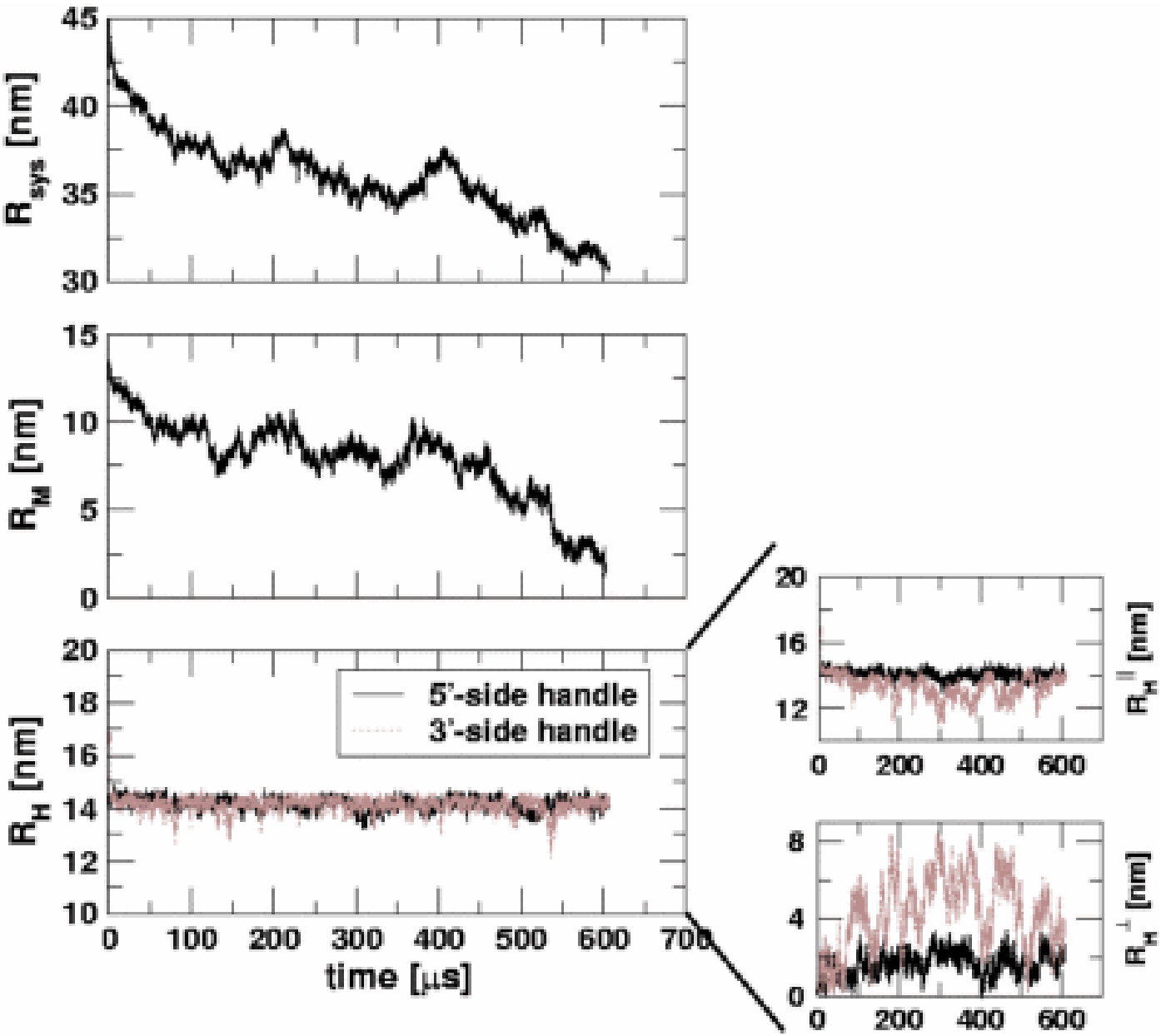}
\caption{\label{refoldlinker}}
\end{figure}

\newpage
\begin{figure}[ht]
\includegraphics[width=6.00in]{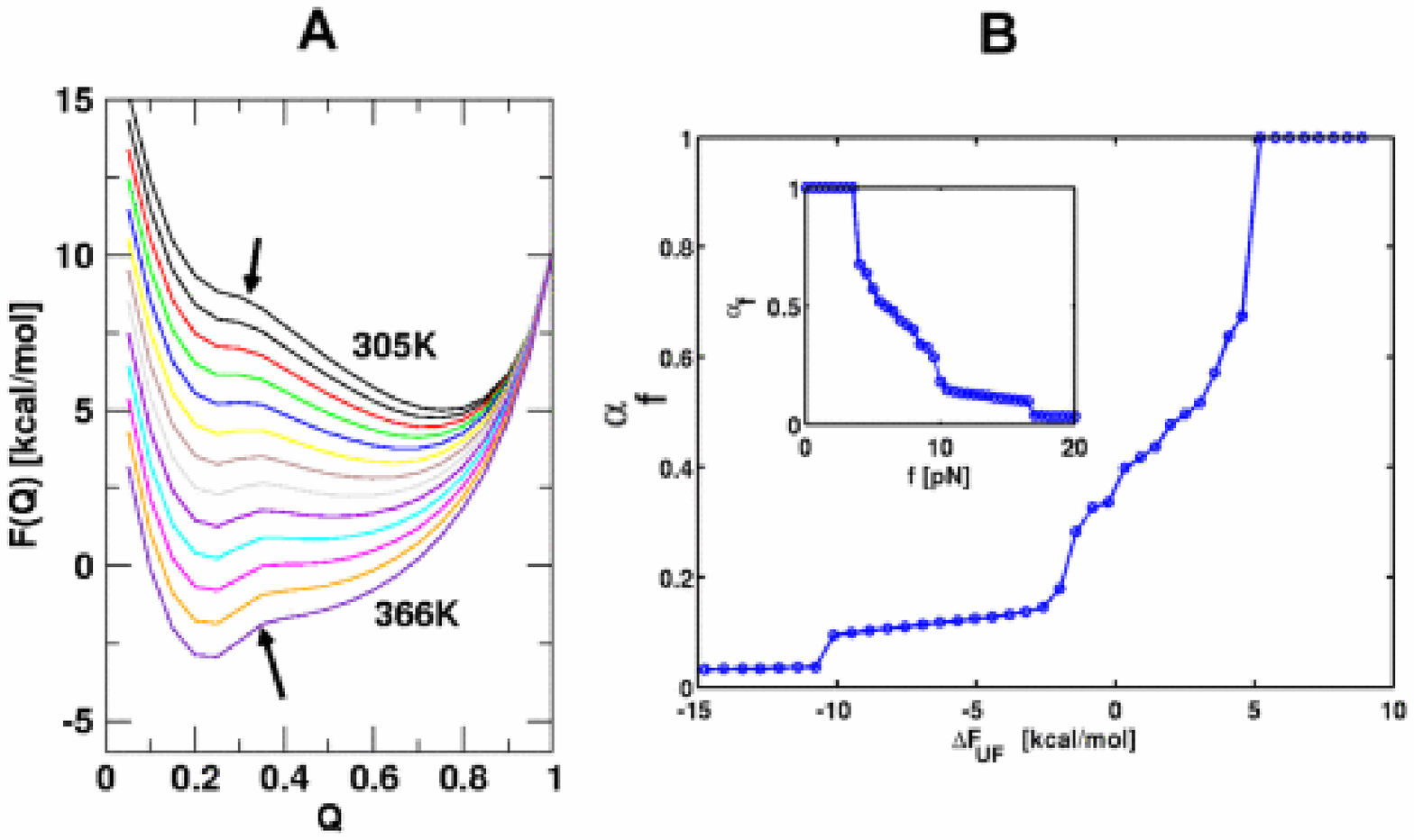}
\caption{\label{Hammond}}
\end{figure}

\newpage
\begin{figure}[ht]
\includegraphics[width=5.00in]{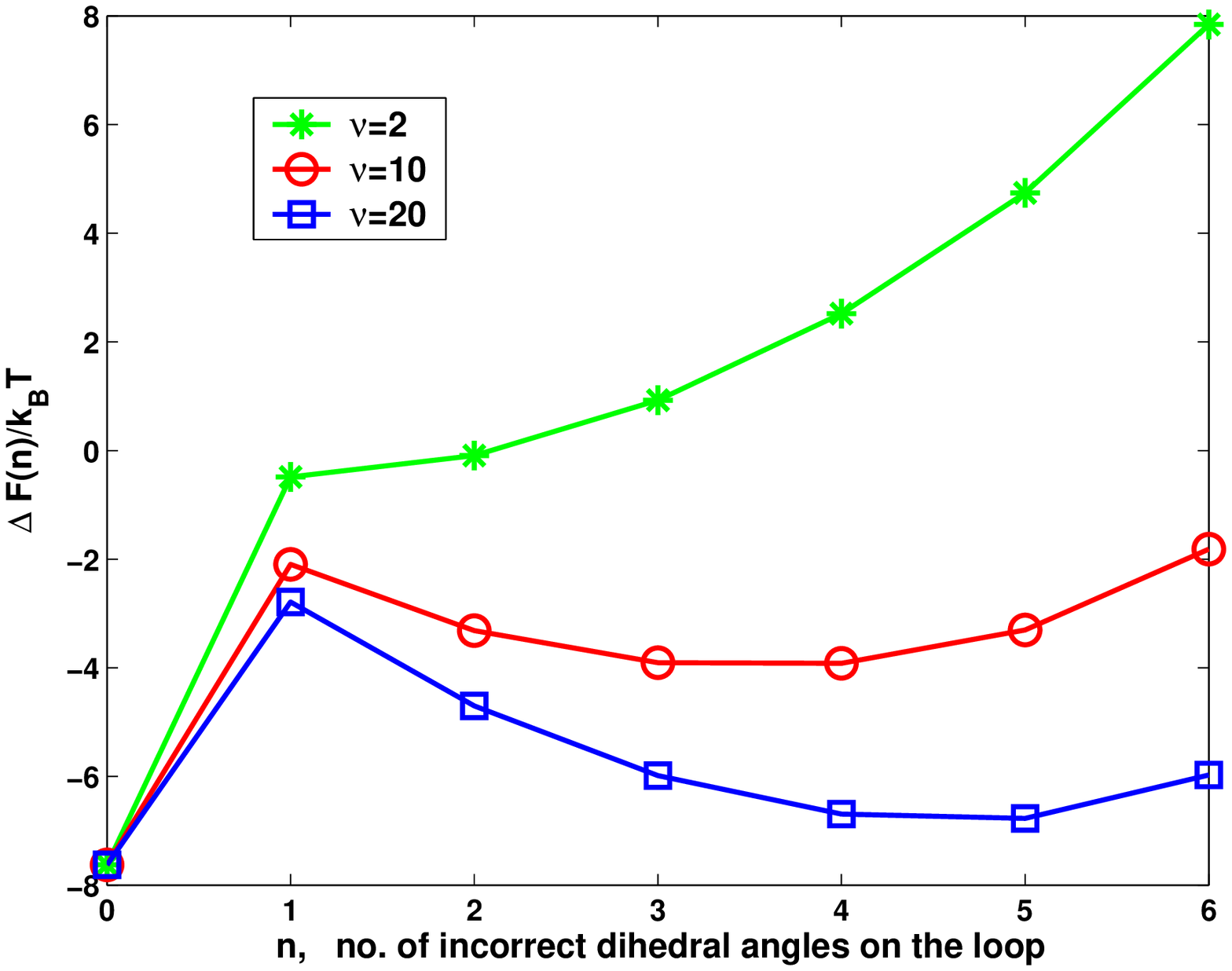}
\caption{\label{Zwanzig}}
\end{figure}

\begin{figure}[ht]
\includegraphics[width=4.00in]{exp_TS_shift.eps}
\caption{\label{exp_TS_shift}}
\end{figure}
\newpage
\begin{figure}[ht]
\includegraphics[width=7.00in]{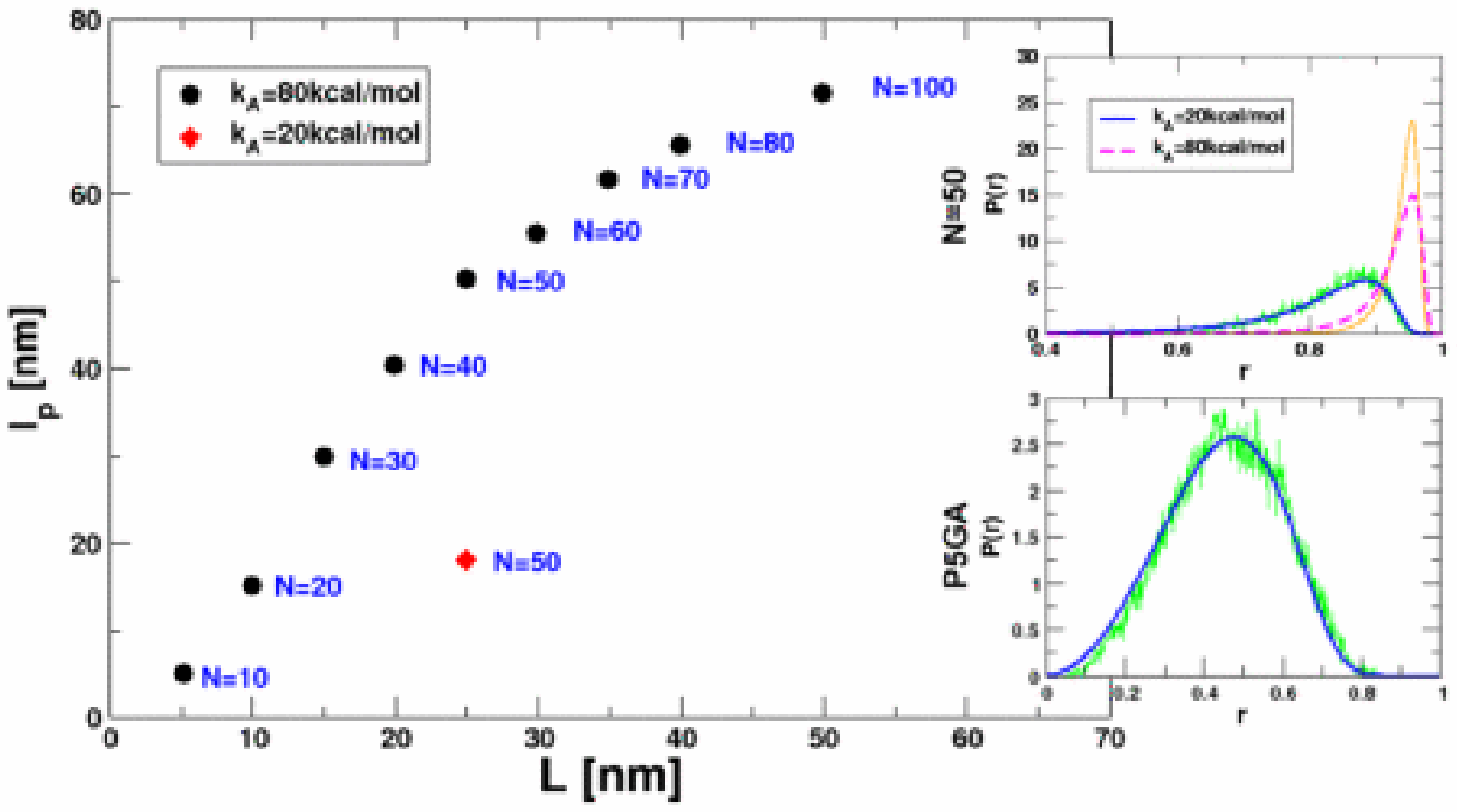}
\caption{\label{lp_comp}}
\end{figure}

\end{document}